\documentclass[manuscript]{acmart}
\usepackage{threeparttable}
\usepackage{xcolor}
\usepackage{arydshln}
\usepackage{multirow}

\AtBeginDocument{%
  \providecommand\BibTeX{{%
    \normalfont B\kern-0.5em{\scshape i\kern-0.25em b}\kern-0.8em\TeX}}}

\setcopyright{acmcopyright}
\copyrightyear{2018}
\acmYear{2018}
\acmDOI{10.1145/1122445.1122456}

\acmConference[Woodstock '18]{}{June 03--05, 2018}{Woodstock, NY}
\acmBooktitle{
  June 03--05, 2018, Woodstock, NY}
\acmPrice{15.00}
\acmISBN{978-1-4503-XXXX-X/18/06}

\begin{document}

\title[Beyond voice assistants]{Beyond Voice Assistants: Exploring Advantages and Risks of an In-Car Social Robot in Real Driving Scenarios}

\author{Yuanchao Li}
\email{yuanchao.li@ed.ac.uk}
\affiliation{
  \institution{University of Edinburgh}
  \country{UK}
}

\author{Lachlan Urquhart}
\email{lachlan.urquhart@ed.ac.uk}
\affiliation{
  \institution{University of Edinburgh}
  \country{UK}
}

\author{Nihan Karatas}
\email{karatas@mirai.nagoya-u.ac.jp}
\affiliation{
  \institution{Nagoya University}
  \country{Japan}
}

\author{Shun Shao}
\email{ss3047@cam.ac.uk}
\affiliation{
  \institution{University of Cambridge}
  \country{UK}
}

\author{Hiroshi Ishiguro}
\email{ishiguro@irl.sys.es.osaka-u.ac.jp}
\affiliation{
  \institution{Osaka University \& ATR}
  \country{Japan}
}

\author{Xun Shen}
\email{shenxun@eei.eng.osaka-u.ac.jp}
\affiliation{
  \institution{Osaka University}
  \country{Japan}
}

\renewcommand{\shortauthors}{Yuanchao, \emph{et al.}}

\begin{abstract}
In-car Voice Assistants (VAs) play an increasingly critical role in automotive user interface design. However, existing VAs primarily perform simple `query-answer' tasks, limiting their ability to sustain drivers' long-term attention. In this study, we investigate the effectiveness of an in-car Robot Assistant (RA) that offers functionalities beyond voice interaction. We aim to answer the question: \emph{How does the presence of a social robot impact user experience in real driving scenarios?} Our study begins with a user survey to understand perspectives on in-car VAs and their influence on driving experiences. We then conduct non-driving and on-road experiments with selected participants to assess user experiences with an RA. Additionally, we conduct subjective ratings to evaluate user perceptions of the RA's personality, which is crucial for robot design. We also explore potential concerns regarding ethical risks. Finally, we provide a comprehensive discussion and recommendations for the future development of in-car RAs.
\end{abstract}



\maketitle

\section{Introduction}
\label{intro}
AI technologies, such as speech recognition, emotion recognition, and dialogue management, have gained popularity in the automotive industry \cite{fukui2018sound,hirsch2000aurora,amman2017impact,li2018towards,shen2020cooperative}. These innovations aim to facilitate human-like interactions for an enhanced driving experience. Integrating such advanced AI capabilities has made the design of in-car voice assistants (VAs) a focal point in recent years. For instance, Mercedes-Benz developed its VA, the \textit{Mercedes-Benz User Experience (MBUX)}\footnote{https://www.la.mercedes-benz.com/en/passengercars/technology/mbux.html}, and Honda introduced the voice-enabled AI conversational assistant -- the \textit{Honda Personal Assistant (HPA)}\footnote{https://global.honda/innovation/CES/2020/honda\_personal\_assistant.html}. These systems enable drivers to intuitively control various functions, including navigation, music, and air conditioning, through voice commands.

Nonetheless, the adoption of in-car VAs has fallen short of expectations due to several reasons. Firstly, the robustness of speech recognition in in-car environments remains a challenge, with speech recognition prone to errors due to noises, unstable network connections, and other passengers' voices \cite{noll1990problems, van1990speech}. Secondly, executing voice commands, such as adjusting the temperature, often takes more time than manual operations (i.e., a simple click or button rotation) \cite{lo2013development}. Thirdly, most assistants are limited to basic `query-answer' tasks, lacking social interaction that could sustain long-term user engagement \cite{li2019expressing}. To address the limitations of existing in-car VAs, the automotive industry has explored the integration of social robots, aiming to create a more enriching driving experience for users \cite{williams2013towards}. However, previous attempts have not seamlessly integrated VA functions and robot interactions into a unified architecture, resulting in Robot Assistants (RAs) often being limited to a robotic appearance without authentic robot behaviors.

In this study, we employ a genuine in-car social robot to explore its impact on enhancing the driving experience. Initially, we conduct a semi-structured interview with 30 participants from diverse backgrounds to explore their perceptions and satisfaction levels of in-car VAs, as well as how these assistants influence their driving experiences. Subsequently, participants are invited to experience an in-car VA equipped with a social robot. We first perform non-driving experiments to familiarize participants with the RA, followed by on-road experiments to evaluate their driving experiences. Moreover, subjective ratings are utilized to assess participants' perceptions of the RA's personality. In addition to the benefits that the RA brings, we also explore its risks by prompting the participants' concerns through another semi-structured interview. Concluding the experiments, we reveal that participants exhibit increased interest in engaging and conversing with the RA. They express feelings of companionship and reassurance, especially during challenging traffic situations. Furthermore, the participants are also concerned about some ethical and legal risks, such as privacy leakage and safety issues. Finally, we provide several recommendations for the future development of in-car RAs and VAs based on our experimental findings. The key contributions of this research are summarized as follows.

\textbf{1)} For the first time, we conduct a comprehensive user study with participants from diverse countries, professions, and age groups, comparing daily-use VAs, traditional in-car VAs, and an in-car RA from multiple perspectives.

\textbf{2)} We comprehensively measure users' feelings and attitudes towards an in-car RA through various quantitative and qualitative metrics. To our knowledge, this is the pioneering study that conducts realistic non-driving and on-road experiments with a real in-car social robot, that is explicitly designed for in-car infotainment purposes rather than employing driving simulators, robot prototypes, or commercially available daily-use robots.

\textbf{3)} We highlight both the benefits and risks of the in-car RA, along with its potential developmental pathways, offering valuable recommendations for both automotive academia and industry.

The remainder of this paper is structured as follows: Section~\ref{sec:work} outlines the existing literature exploring various aspects of daily-used and in-car assistants. Following this, Section~\ref{sec:method} describes the robot and methodology employed in this study. Section~\ref{sec:interview} introduces the user study, its analysis results, and ensuing discussions. Sections~\ref{sec:evaluation} and \ref{sec:personality} detail the experimental evaluation, including non-driving and on-road experiments, as well as personality assessment, highlighting the advantages of the in-car RA. Section~\ref{sec:concerns} investigates its potential risks through an interview study. Subsequently, we conclude the evaluation with discussions in Section~\ref{sec:discussion} and provide relevant recommendations for the community in Section~\ref{sec:recommendations}. Finally, we summarize this study in Section~\ref{sec:conclusion}.

\section{Related Work}
\label{sec:work}

\subsection{Daily-Used Voice Assistants}
Voice assistants, also known as speech interfaces and smart speakers, have significantly developed over the past decade. Initially popular with \textit{Apple Siri}\footnote{https://www.apple.com/siri}, \textit{Amazon Alexa}\footnote{https://developer.amazon.com/en-US/alexa}, \textit{Google Home}\footnote{https://assistant.google.com}, and others have emerged, offering expanded functionalities beyond controlling smartphones to managing various intelligent devices such as home appliances \cite{nafari2013augmenting,portet2013design}. Voice operations hold advantages over manual tasks in several ways: 1) They are more direct and time-efficient, often eliminating several intermediate manual steps. 2) They are less confined by spatial limitations, allowing the execution of commands as long as the voice signals reach the device. An important prerequisite for these VAs to work smoothly is that voice commands from human users are accurately recognized.

\subsection{In-Car Voice Assistants}
VAs are now integral to modern in-car systems, handling various tasks such as controlling radios and music players, making phone calls, and managing navigation via simple voice commands \cite{lo2013development,tashev2009commute}. These assistants have demonstrated their effectiveness in different scenarios. For instance, in elderly driving, data collected during interactions with in-car VAs has been used to predict potential accident risks by identifying subtle cognitive impairments \cite{yamada2021using}. Additionally, in general driving situations, proactive voice assistants have reduced drivers' cognitive load and increased their satisfaction \cite{schmidt2020users}. In today's in-car systems, VAs primarily fall into two categories. The first is an extension of everyday VAs, such as \textit{Google Android Auto}\footnote{https://www.android.com/auto} and \textit{Apple CarPlay}\footnote{https://support.apple.com/guide/iphone/use-siri-in-your-car-iphadacf963f/ios}, which are specifically adapted to be used in cars. We refer to this category as \textit{phone-based} since it requires a Bluetooth connection or a specific App link to the car. The second category comprises in-car systems themselves, like \textit{MBUX}, which do not require phone connectivity and can control car-specific functions. We term this category \textit{car-based}.

However, various noises encountered during driving, including passenger conversations, traffic sounds like honking and passing vehicles, wind noise at higher speeds, and engine sounds, often render in-car VAs more error-prone compared to their daily-use counterparts \cite{horswill2008auditory,kim2000psychophysiological,li2018towards}. Furthermore, complex and changing driving environments, such as rural areas or underground tunnels, can disrupt network signals, leading to delays or downtime for online VAs \cite{lauridsen2017lte}. Consequently, these issues might diminish drivers' willingness to use in-car VAs. To this end, researchers, engineers, and UI/UX designers are exploring diverse solutions, including the utilization of RAs to mitigate these problems and enhance the overall driving experience.

\subsection{In-Car Robot Assistants}
Several car manufacturers have ventured into the use of RAs, such as \textit{BYD Qin}\footnote{https://www.theverge.com/2012/4/23/2968502/byd-qin-robot-beijing-auto-show-aida-robotic-assistant}, \textit{Nissan PIVO}\footnote{https://global.nissannews.com/en/releases/release-745329929e3407eb4c295c25eb02d5a1/photos/17}, and \textit{Pioneer Carnaby}\footnote{https://techcrunch.com/2010/05/20/carnaby-pioneers-very-special-car-navigation-mini-robot-videos}, to introduce features like music, radio, navigation, and even fatigue detection. However, most of these robots remain in the concept car stage and have not gained popularity in everyday driving scenarios. According to Williams \emph{et al.} \cite{williams2013towards}, these robots separate functionality and social interaction, whereas it is essential to integrate these aspects into a seamless architecture. To address this, they developed an Affective Intelligent Driving Agent (AIDA), an in-car companion robot utilizing the driver's smartphone as its face, providing information about the phone, the vehicle, and the city environment to the driver. Experimental evaluations showed that AIDA could better assist drivers and enhance sociability compared to traditional smartphones during simulated driving tasks. Moreover, Karatas \emph{et al.} \cite{karatas2016namida} introduced NAMIDA, a robot with a context-aware interaction for multi-party conversations during simulated driving. Their experimental analysis revealed NAMIDA's potential in reducing system-induced attention overload. Additionally, Wang \emph{et al.} \cite{wang2021vehicle} adopted robot Nao in a simulated fully autonomous driving scenario, exploring the effects of speech style and embodiment together and separately. Their experimental results showed the capability of robot agents in promoting drivers' likability and perceived warmth, as well as receiving higher competence and lower perceived workload scores.

\begin{figure}[ht]
  \centering
  \includegraphics[scale=0.25]{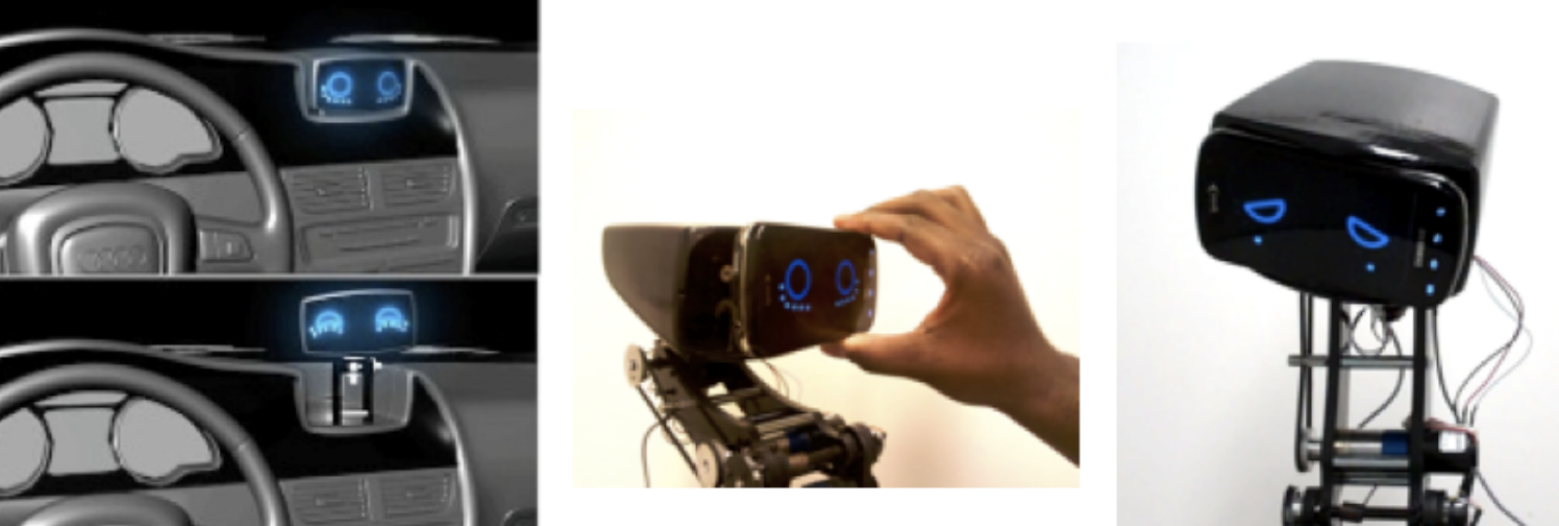}
  \hspace{8pt}
  \includegraphics[scale=0.25]{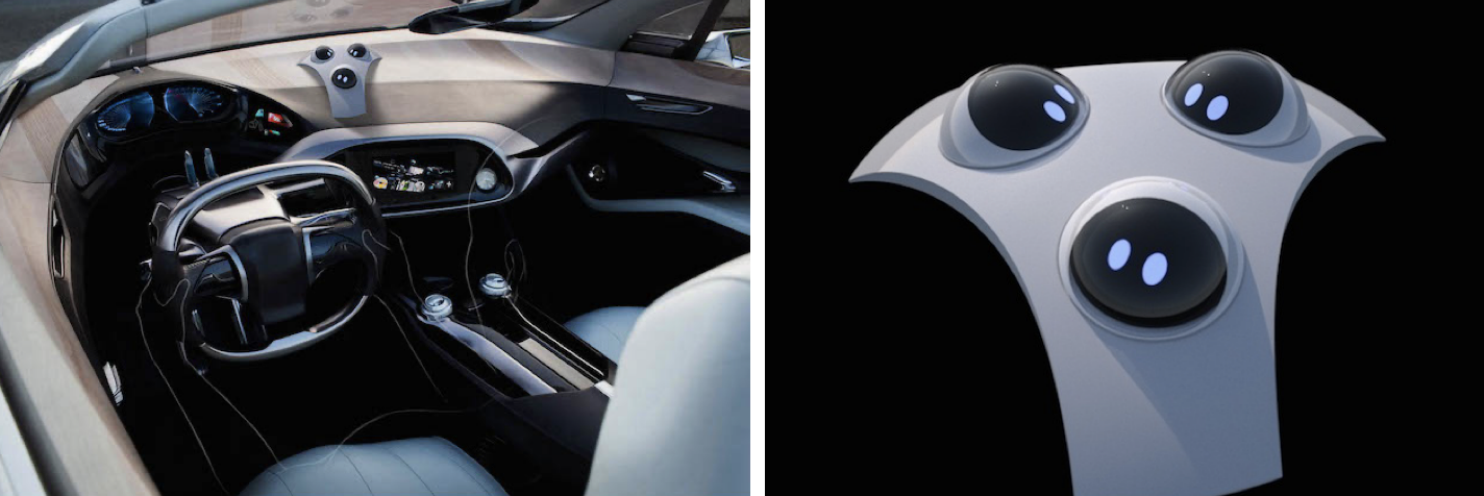}
  \caption{AIDA (three on the left) and NAMIDA (two on the right).}
  \label{fig:robots}
\end{figure}

The study conducted by Tanaka \emph{et al.} \cite{tanaka2020analysis} demonstrated that integrating a small humanoid robot (\textit{Robohon}, SHARP Co., Ltd.) on the dashboard positively influenced driving operations. Results indicated high acceptability of the agent, minimal driver distraction in the actual car environment, and an improvement in driving behavior. Long-term experiments with Robohon further underscored its positive impact on mitigating risky driving behaviors in elderly individuals, such as reducing speeds when entering intersections \cite{tanaka20234}. However, humanoid or highly anthropomorphized robotic assistants tend to raise the expectations of their human interlocutors regarding their interaction capabilities. Research has shown that when users' perceptions of robots fall short of their expectations, they may feel disappointed and skeptical of the robot's outputs, which results in a high adaptation gap \cite{komatsu2012does}. Consequently, engaging with humanoid or highly anthropomorphic RAs within a driving environment may evoke similar feelings of dissatisfaction in their interactions with human counterparts, as their human-like appearance and behavior may lead to elevated expectations. Therefore, it is important to consider the level of anthropomorphism in RAs.

Zihsler \emph{et al.} focused on improving trust in self-driving cars by integrating human-like behaviors and expressions into RAs resembling a human-like head \cite{zihsler2016carvatar}. Also, Cheng \emph{et al.} \cite{cheng2022driving} explored the impact of anthropomorphic design of the RA on trust and driving performance, finding that users familiar with the in-car infotainment system trusted highly anthropomorphized RA less compared to unfamiliar users. Moreover, Karatas \emph{et al.} \cite{karatas2023robotic} emphasized the significance of minimal design, excluding non-essential anthropomorphic elements, and considering their expressiveness for the acceptance of RAs in driving situations.

Despite the existence of the aforementioned studies, the RAs utilized in these studies either originated from mobile phones, external robots, or were in the prototype research phase. None of them are currently integrated into car systems as a real robot. In the context of in-car environments, the design of the RA should seamlessly integrate with the driving environment. Therefore, in this work, we employ a genuine physical social robot, which is integrated with the VA function, to explore drivers' real-time experiences and perceptions while driving an actual production car. Our goal is to conduct the very first investigation of a seamless in-car RA, providing timely and valuable insights to existing literature and relevant communities.

\section{Methodology}
\label{sec:method}

\subsection{Social robot NOMI}
\begin{figure}[ht]
  \centering
  \includegraphics[scale=0.11]{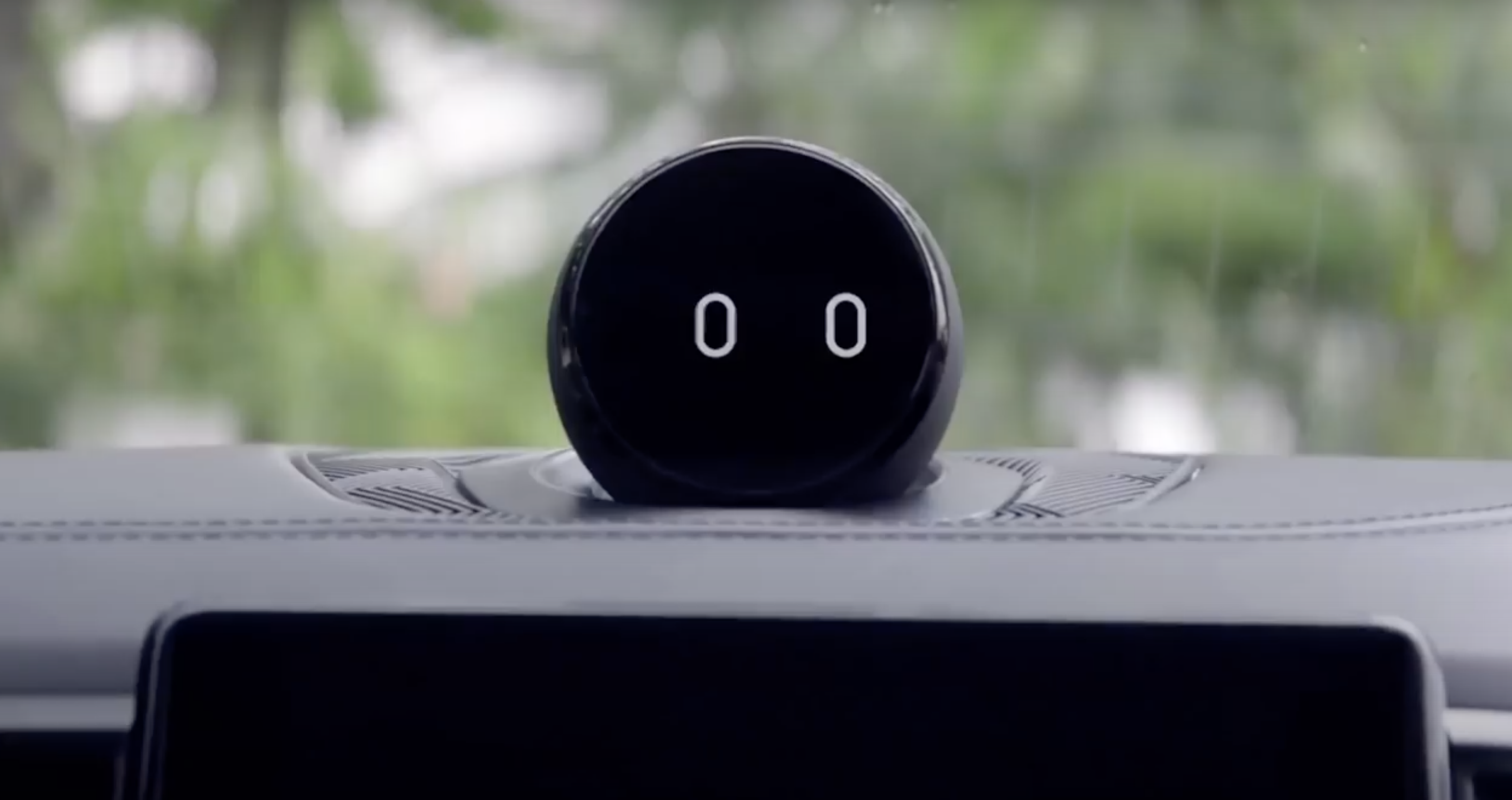}
  \hspace{8pt}
  \includegraphics[scale=0.14]{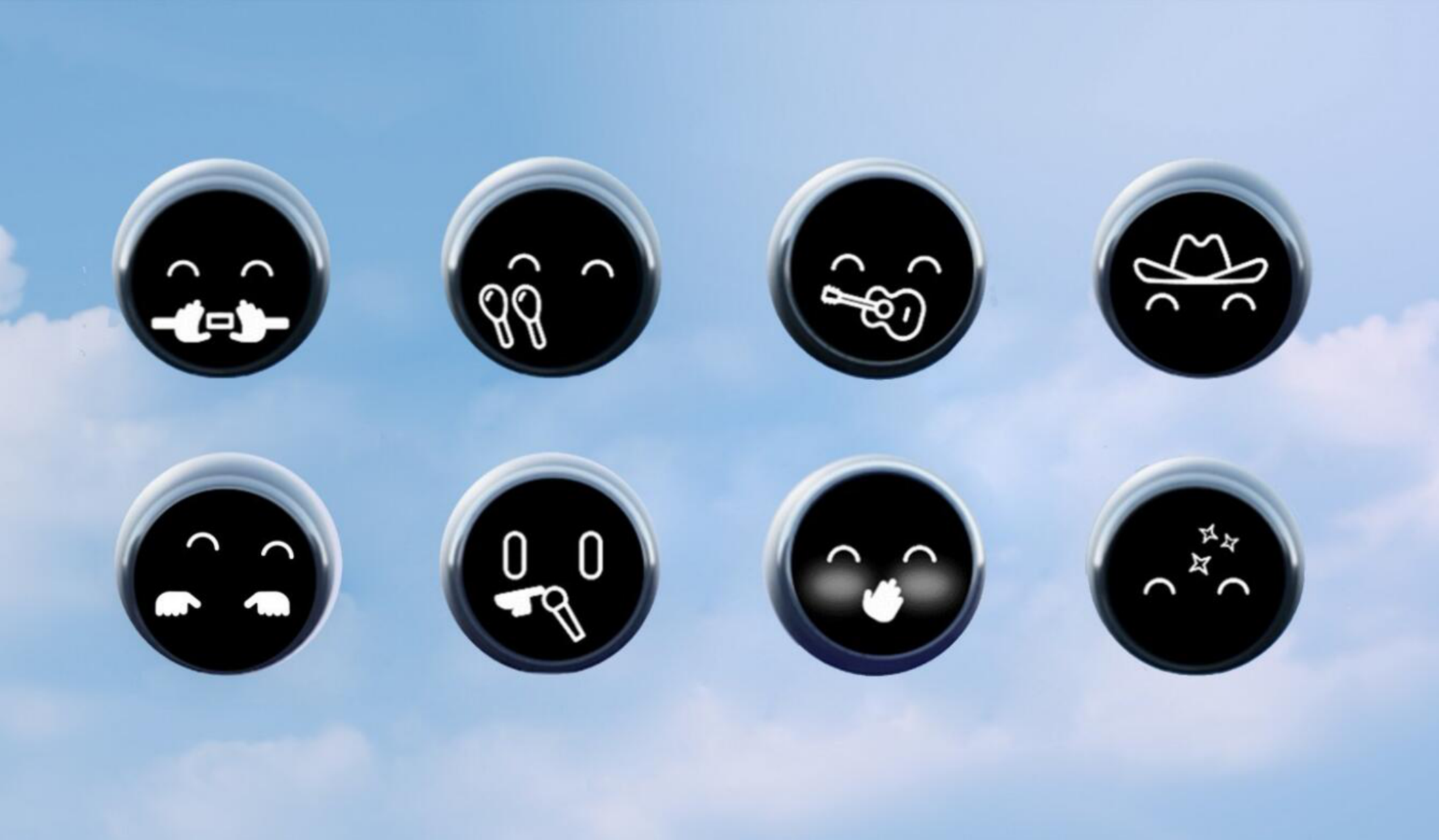}
  \caption{Social robot NOMI (left) and some of its facial expressions (right).}
  \label{fig:nomi}
\end{figure}

In this study, we utilize \textit{NOMI}, an AI companion developed by the electric vehicle company NIO\footnote{https://www.nio.com}. NOMI is a small ball-shaped dashtop device designed in the form of a head and face, facing inward toward the inside of the car. NOMI responds to voice commands by rotating within a defined angle toward the speaker and can display various facial expressions to convey feelings, weather conditions, safety alerts, music suggestions, and self-expressions (see Fig.~\ref{fig:nomi}). In most cases, NOMI tends to exhibit a cheerful demeanor, creating a relaxed atmosphere. It functions as an AI companion and social robot, which enhances the VA system, aiming to comprehend the driver's needs. Its name, pronounced similar to `Know Me,' reflects its purpose.

\begin{figure}[ht]
  \centering
  \includegraphics[scale=0.55]{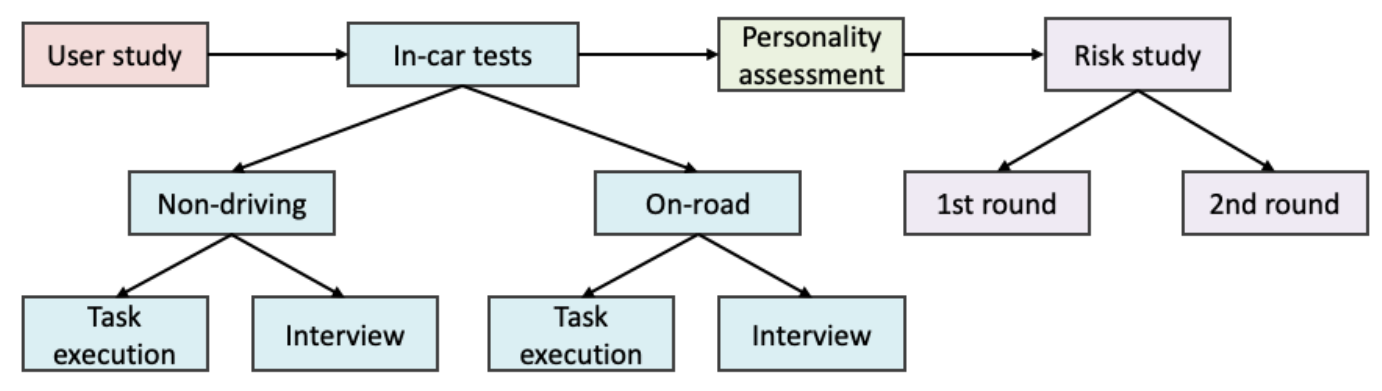}
  \caption{Key steps of the research approach.}
  \label{fig:step}
\end{figure}

\subsection{Research Approach}
We undertake four key steps to achieve our objectives (see Fig.~\ref{fig:step}). Firstly, we conduct an user study (Sec.~\ref{sec:interview}) by performing a semi-structured interview with participants to assess their experience and familiarity with VAs. Next, we carry out in-car tests (Sec.~\ref{sec:evaluation}), engaging participants in conversation and tasks with NOMI across diverse driving scenarios. Participants then rate their driving experiences and provide feedback for our analysis. In both non-driving and on-road experiments, participants undergo the tasks twice: once without the presence of NOMI robot (i.e., visual access to NOMI), relying solely on voice interaction, and then with both voice and visual access. This aims to explore the distinctions between the in-car VA's performance with and without the presence of a social robot. Subsequently, participants evaluate NOMI's perceived personality, comparing to their prior VA experiences (Sec.~\ref{sec:personality}). Finally, we conduct a risk study (Sec.~\ref{sec:concerns}) through a semi-structured interview with the participants. The experiments involve both quantitative and qualitative analyses, detailed in the following sections.

\section{User Study}
\label{sec:interview}

\subsection{Survey and Interviews}
\begin{figure}[ht]
  \centering
  \includegraphics[width=0.85\textwidth]{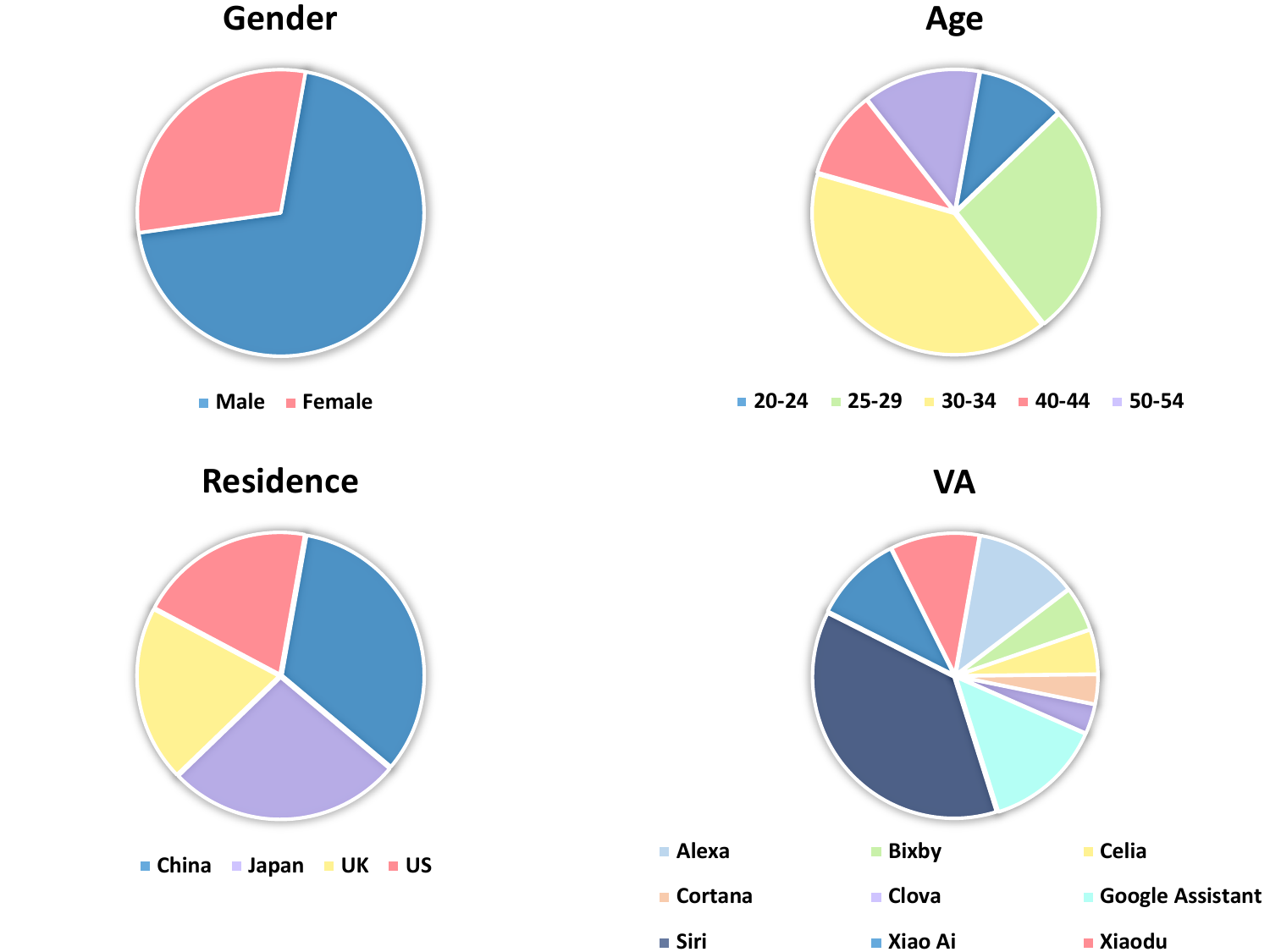}
  \caption{Ratio visualization of participants' demographic information.}
  \label{fig:pie}
\end{figure}

First, we conduct semi-structured interviews with 30 participants (21 males and 9 females, ages ranging from 25 to 55) who have experience using in-car VAs (either phone-based or car-based). Before the interviews, the participants are first required to fill in a survey of their demographic information, along with the VAs they have used (not including in-car VAs), which are shown in Table~\ref{tab:profile}. To ensure that the results of the user study are generalizable, we recruit participants of both genders, ranging in age from twenty to fifty generations, covering a wide range of professions, and living in several different countries. We present pie charts based on these aspects for clear visualization in Fig.~\ref{fig:pie}. From the participants' demographics, we can obtain some interesting findings: \textbf{1)} \textit{Apple Siri} is the most commonly used one among all VAs. \textbf{2)} Users in China are more likely to use local brands likely because some foreign brands are not available whereas users in Japan are less likely to use local brands. \textbf{3)} Younger and male users are more likely to have experience of using more VAs. These findings may be specific to this survey, but they could also reflect the attitudes and usage habits of users from different groups, thereby aiding in the development of suitable in-car VAs accordingly.

\begin{table}[ht]
\centering
    \caption{Summary of participants' demographic information.}
    \label{tab:profile}
    \begin{threeparttable}
    \begin{tabular}{llllll}
    \toprule
    \textbf{Participant} & \textbf{Gender} & \textbf{Age} & \textbf{Profession} & \textbf{Residence} & \textbf{VA\tnote{1}}\\
    \midrule
    P1 & Male & 30-34 & Designer & Japan & Clova, Siri \\
    P2 & Male & 30-34 & Civil Engineer & China & Siri, Xiaodu, Xiao AI \\
    P3 & Female & 25-29 & Research Fellow & US & Alexa, Google Assistant \\
    P4 & Male & 25-29 & System Engineer & Japan & Clova, Siri \\
    P5 & Male & 25-29 & Software Engineer & China & Xiaodu, Xiao AI \\
    P6 & Female & 25-29 & Accountant & China & Xiao AI \\
    P7 & Male & 25-29 & Graduate Student & China & Siri, Xiaodu \\
    P8 & Female & 30-34 & Self-Employed & Japan & Alexa, Siri \\
    P9 & Female & 30-34 & HR Manager & US & Bixby, Siri \\
    P10 & Male & 30-34 & University Faculty & Japan & Alexa, Google Assistant, Siri \\
    P11 & Male & 30-34 & Sales Specialist & Japan & Siri \\
    P12 & Male & 50-54 & Software Engineer & Japan & Google Assistant, Siri \\
    P13 & Male & 50-54 & Consultant & China & Celia, Xiaodu \\
    P14 & Male & 50-54 & Sales Specialist & China & Xiao AI \\
    P15 & Male & 40-44 & Self-Employed & UK & Siri \\
    P16 & Male & 30-34 & Research Fellow & UK & Google Assistant, Siri \\
    P17 & Female & 30-34 & Graduate Student & UK & Bixby, Siri \\
    P18 & Male & 30-34 & Software Engineer & US & Google Assistant, Siri \\
    P19 & Female & 30-34 & Primary School Teacher & China & Siri, Xiaodu, Xiao AI \\
    P20 & Male & 25-29 & Research Fellow & China & Xiaodu, Xiao AI \\
    P21 & Male & 25-29 & Graduate Student & Japan & Siri \\
    P22 & Female & 40-44 & Sales Specialist & Japan & Google Assistant \\
    P23 & Male & 40-44 & Project Manager & China & Celia, Siri \\
    P24 & Male & 50-54 & Self-Employed & China & Celia \\ 
    P25 & Male & 20-24 & Graduate Student & UK & Google Assistant, Siri \\ 
    P26 & Male & 20-24 & Graduate Student & UK & Cortana, Siri \\ 
    P27 & Female & 30-34 & University Faculty & UK & Alexa, Siri \\ 
    P28 & Female & 20-24 & Graduate Student & US & Alexa, Google Assistant, Siri \\ 
    P29 & Male & 25-29 & Electronics Engineer & US & Alexa, Cortana, Siri \\ 
    P30 & Male & 30-34 & Electronics Engineer & US & Alexa, Bixby, Siri \\ 
    \bottomrule
    \end{tabular}
    \begin{tablenotes}
    \footnotesize
    \item[1]Alexa (by Amazon); Bixby (by Samsung); Celia (by Huawei); Clova (by Line); Cortana (by Microsoft); Google Assistant (by Google); Siri (by Apple); Xiaodu (by Baidu); Xiao AI (by Xiaomi).
    \end{tablenotes}
    \end{threeparttable}
\end{table}

Subsequently, participants are asked to engage in one-to-one interviews comprising general questions. All the interviews are conducted remotely over LINE, Messenger, WeChat, or Zoom. We begin the interviews with a brief greeting and a discussion related to the participants' survey information before delving into more detailed interview questions. The interviews last on average 12 minutes, ranging from 10 to 15 minutes. Prior to the interviews, we inform all participants that the interview will be recorded, and they retain the right to request data deletion.

The interview questions are structured into four parts. Part I discusses the general usage of VAs by participants. We inquire about basic information, such as the brands they have used and the purposes for which they utilize these VAs. In Part II, our focus shifts to the participants' attitudes toward in-car VAs. For instance, participants are asked about how the assistants affect their driving experience and their feelings about in-car VAs, especially in comparison to their previously or regularly used daily VAs. Part III involves participants sharing their expectations for in-car VAs, including functions they deem unnecessary and additional features they would like to see in future assistants.

The complete set of interview questions is listed in Table~\ref{tab:interview}. It is worth noting that, depending on participants' answers, we occasionally offer feedback or pose follow-up questions to ensure comprehensive answers. These spontaneous feedback and follow-up questions are not included in the table as they do not influence our designed questions or the resulting answers.

\begin{table*}[ht]
  \caption{Interview questions for the survey.}
  \label{tab:interview}
  \begin{center}
  \begin{tabular}{l}
    \toprule
    \textit{\textbf{Part I}} \\
    \hdashline
    Q1. When did you start using VAs, and what were the reasons? \\
    Q2. Which in-car assistants have you used, and what functions do you utilize? \\
    \midrule
    \textit{\textbf{Part II}} \\
    \hdashline
    Q3. Have the in-car assistants enhanced your driving experience? \\ \ \ \ \ \ \ \ If yes, in what ways? If not, what are the reasons? \\
    Q4. How do you perceive in-car assistants compared to the VAs you use on a daily basis? \\
    \midrule
    \textit{\textbf{Part III}} \\
    \hdashline
    Q5. What specific aspects of the in-car assistants are you least satisfied with? \\ \ \ \ \ \ \ \ And what are the reasons behind this dissatisfaction? \\
    Q6. How would you rate your satisfaction with the current in-car assistants? \\
    \bottomrule
  \end{tabular}
  \end{center}
\end{table*}

\subsection{Analysis Results}
We present overall conclusions, highlighting selected representative or exceptional answers as examples.

\textbf{Part I - Q1}.
The majority of participants began using VAs out of curiosity about AI, likely influenced by advertisements a decade ago. Their experience with VAs spans no more than seven years, except for those familiar with using \textit{Siri}.

\textit{``AlphaGo was my first exposure to artificial intelligence, and after that, all sorts of AI-based products started popping up. I ended up buying one just to see how smart it was, probably around 2017.'' (P3)}

\textit{``Over 10 years ago, I began using Siri. As a huge Apple fan, it just felt natural to incorporate Siri into my routine.'' (P11)}

\textit{``I bought this phone just last year, and it had this thing (i.e., the VA) already there, so I started using it. Sometimes it just pops up on its own. I'm not really into gadgets, so I don't know how to shut this thing off.'' (P24)}

We can observe that most of them started using VAs due to the emergence of new technological products, but certain participants feel somewhat `obligated' to use VAs.

\textbf{Part I - Q2}.
To avoid explicit comments about specific brands, especially negative ones, we refrain from indicating which participants used which brands. Among the in-car assistants used in this survey are \textit{Skoda Laura\footnote{https://www.skoda-storyboard.com/en/skoda-world/innovation-and-technology/okay-laura-lets-talk}}, \textit{BMW IPA\footnote{https://www.bmw.co.uk/en/topics/owners/bmw-connecteddrive/intelligent-personal-assistant.html}}, \textit{Mercedes MBUX}, \textit{Apple CarPlay}, \textit{Google Android Auto}. Other participants' assistants are unnamed, so we only mention the car models: \textit{Chevrolet Monza 330}, \textit{Geely Emgrand L}, \textit{ORA Good Cat}, \textit{Tesla Model 3}, \textit{Tesla Model Y}, and \textit{XPENG P7}.

The music function is used by all participants, which is unsurprising. However, the usage of other functions differs significantly among the assistant categories. Since phone-based users lack the ability to control the car via voice commands, phone calls are the second most frequently used function among these participants. In contrast, car-based users show a preference for operating car functions, such as temperature adjustment and window controls. Nevertheless, both categories of users frequently use navigation, except for participants in China.

\textit{``I usually take calls and listen to music. (Follow-up question: What about other functions? For example, navigation.) I use the phone for navigation. I open the navigation routes and then place the phone in a stand attached to the air conditioning vent, allowing me to keep an eye on the map and other applications.'' (P2)}

\textit{``The map application on the phone is regularly updated, but the version in the car doesn't seem to update in real time. I've been misled by the car navigation function a couple of times.'' (P5)}

\textit{``I struggle with the map application because it's different from the one I use on my phone.'' (P6)}

It is evident that Chinese users heavily rely on their phones, expecting in-car assistants to adapt to their existing usage habits rather than learning entirely new assistant features. If the assistants do not integrate well with phone applications, these users might opt to abandon some of the assistant's functions.

\textbf{Part II - Q3}.
Participants from the US, UK, and China commonly agree that in-car assistants enhance their driving experiences. However, this sentiment does not align with Japanese participants. In the US, all participants expressed their enjoyment of using in-car assistants, especially during long-distance drives. We hypothesize that long hours of driving might lead to driver fatigue, and the presence of in-car assistants could alleviate boredom and add enjoyment to the driving experience.

\textit{``Yeah, I absolutely love using the in-car VA. See, I've got this daily trek to school that takes up almost an hour each way. It's this road surrounded by farmland, not many cars around, so, you know, it's a bit of a snooze-fest driving through. I just plug in some tunes to pass the time, and using voice commands to switch tracks is super handy. It totally jazzes up my driving experience.'' (P3)}

\textit{``I feel there's been an improvement of the driving experience. My work often involves interacting with customers, so I'm frequently on the phone in my car, navigating to appointments. It's really handy to use voice commands for these tasks while driving. Plus, on weekends, I often take the kids out, and you know how kids are, they love chatting with the VA.'' (P9)}

Participants in China provided similar feedback, noting that in-car VAs alleviate driving boredom and they enjoyed controlling car functions via voice, perhaps because the most popular in-car VAs in China are car-based rather than phone-based. However, many participants in Japan express a contrasting attitude to those in other countries, opining that in-car VAs are not particularly necessary for Japan.

\textit{``It's hard to say. I rarely drive because Japan's train system is highly developed. And even when I do drive, the distances aren't long. So, I don't feel that in-car VAs are particularly necessary.'' (P10)}

\textit{``I don't notice much difference. I don't usually listen to music while driving. Besides, many companies in Japan manufacture and sell specialized car navigation devices, which are quite popular. Elderly individuals might not opt to use VAs. I'm not entirely sure, perhaps having them is better than not.'' (P12)}

\textit{``I only rent a car for outings with friends during holidays, and since we chat all the time in the car, I don't find many opportunities to use in-car VAs. But when I have my own car, I believe I'll use it a lot.'' (P21)}

Based on these responses, it can be noted that Japan's distinct circumstances, such as the well-established train system, result in less necessity for long driving distances. Consequently, the demand for in-car VAs differs from participants in other countries.

\textbf{Part II - Q4}.
Despite varying attitudes toward in-car VAs among participants from different countries, their responses to this question surprised us. Across all countries, most participants mentioned the need to speak more cautiously while interacting with in-car VAs due to concerns about inaccurate speech recognition or the lack of support for specific functions.

\textit{``When you finish giving a command and wait for the VA to execute it, but it responds by saying it didn't understand, it can be really annoying.'' (P1)}

\textit{``At times, I prefer manual operation rather than speaking to the in-car VA. For instance, when adjusting the air conditioning, I've already completed the action by the time it begins to respond.'' (P5)}

\textit{``There are so many voice commands, and it's practically impossible to learn them all from the manual. I tend to say what naturally comes to mind, but sometimes, the VA responds by saying it doesn't support that function.'' (P15)}

\textit{``Given my work in the AI industry, I've been using VAs for a while, which has shaped my thinking and conversation habits with them. It's almost like I've been `conditioned' to interact this way. But for my father's generation, who lack experience in using voice commands but suddenly have to use them, it can be uncomfortable, particularly because machines can't grasp ambiguous or vague intentions like humans can.'' (P18)}

This common experience underscores the global relevance of challenges with VAs in cars, crossing language and cultural boundaries. Users' shared frustration when VAs struggle to understand or perform commands, as highlighted by P1's irritation, is a sentiment widely echoed. Moreover, P5's preference for manual operations in specific situations emphasizes the necessity for enhancing the reliability of voice commands.

\textbf{Part III - Q5}.
Some participants mention that the functionality of in-car VAs is too limited, especially for those using phone-based ones. Additionally, poor speech recognition performance is also mentioned by some participants.

\textit{``At times, I feel the VA acts as a mere intermediary to the phone, handling tasks like making calls, playing music, checking the weather, and more. Truth be told, all these functions can be executed by operating the phone using voice commands.'' (P2)}

\textit{``The VAs can only execute voice commands, lacking the ability to interact with the driver. It would be remarkable if VAs could resemble those depicted in sci-fi movies, with interactive and engaging capabilities.'' (P30)}

\textit{``Given my major in speech recognition, I believe the speech recognition function should primarily adapt to driving scenarios. When my friends were chatting in the car, the VA couldn't recognize my speech well.'' (P29)}

The feedback from the participants emphasizes the necessity for ongoing innovation in in-car VA systems. These observations serve as guidance for developing VAs that encompass advanced features, human-like interactions, and robust speech recognition specifically tailored to the distinct challenges of the automotive environment.

\textbf{Part III - Q6}.
As anticipated, most participants find the current in-car VAs satisfactory in meeting their needs. However, these VAs do not exceed the participants' expectations.

\textit{``Not bad. When I initially began using it, it felt rather novel. However, I'm not particularly inclined to use it now unless more interesting functions have been developed.'' (P9)}

\textit{``Overall, I'm satisfied. There are occasions when I want to listen to music or when stuck in traffic with nothing to do, and having the VA proves beneficial. On a scale of ten, I'd rate it seven.'' (P19)}

\textit{``As I mentioned, I'd be more satisfied if the speech recognition function could improve. However, I recognize that the current issue stems from limitations in the overall field of speech recognition.'' (P29)}

Building upon the feedback and findings from these questions, it is clear that despite certain unavoidable limitations, the overall experience is satisfactory. However, it appears that the usability is mainly confined to entertainment functions, like playing music. Hence, future development should prioritize improving functionality. Moreover, current in-car VAs seem unable to capture the participants' long-term attention, which requires further investigation into the factors influencing user engagement and strategies to enhance it. This aspect is crucial for designing in-car VAs that can consistently provide value and maintain user interest over extended periods.

\section{In-Car Experiments}
\label{sec:evaluation}
In light of the user study results, we design non-driving and on-road experiments involving the in-car robot NOMI to execute commands from drivers and engage in chatting functions. As NIO cars have not yet entered the markets of the UK, US, and Japan, only ten participants in China could participate in the experiments. However, we believe that the findings should be applicable across different countries, given that NIO has conducted user studies on the design of NOMI by a global team \footnote{https://www.nio.com/blog/history-behind-nomi}. We plan to extend the experiments after NIO releases cars in the other countries.

\subsection{Experiment Tasks}
\label{sec:task}
The experiments comprise the following tasks, all executed solely through the participants' voice commands.

\textbf{\textit{1. Wake-up word}}: At the beginning of the experiment, participants would initiate the conversation by uttering the wake-up phrase ``Hi, NOMI.''

\textbf{\textit{2. POI search}}: Points Of Interest (POI) are specific locations that can be integrated into the in-car system, easily retrieved and displayed on the screen through manual operation or voice command. The designated locations in this task include \textit{charging stations}, \textit{restaurants}, and \textit{parking lots}. Participants are encouraged to convey their desire to locate these POIs using any phrasing they choose, without being constrained to specific commands. For instance, participants could say "I'm hungry" instead of "Find a restaurant."

\textbf{\textit{3. Navigation}}: This task follows task 2 after the POIs are provided by the system. In the non-driving experiment, participants can freely select from the provided candidates. However, in the on-road experiment, participants are mandated to select the first candidate that aligns with our predetermined driving route.

\textbf{\textit{4. AC control}}: Participants are instructed to utilize voice commands to both activate and deactivate the air conditioning, as well as to adjust the temperature and air speed once for each setting.

\textbf{\textit{5. Touchscreen function control}}: Participants need to perform three functions: adjusting screen brightness, enabling and disabling Bluetooth \& WiFi, and returning to the home screen from any other opened function.

\textbf{\textit{6. Seat control}}: Participants are tasked with executing two functions: managing seat ventilation and heating, and controlling airflow and temperature.

\textbf{\textit{7. Window control}}: Participants are instructed to open or close the window individually for the driver seat, passenger seat, left rear seat, right rear seat, and for the entire vehicle.

\textbf{\textit{8. Dialing}}: Participants need to make a call to the first contact and answer their incoming phone call.

\textbf{\textit{9. Entertainment (radio, music, and video)}}: Participants are instructed to activate the radio, music, and video, and then operate their associated controls, such as changing channels, songs, or episodes, and adjusting the volume.

\textbf{\textit{10. Information inquiry}}: Participants need to inquire about the weather and traffic conditions.

\begin{figure*}[ht]
  \centering
  \includegraphics[scale=0.29]{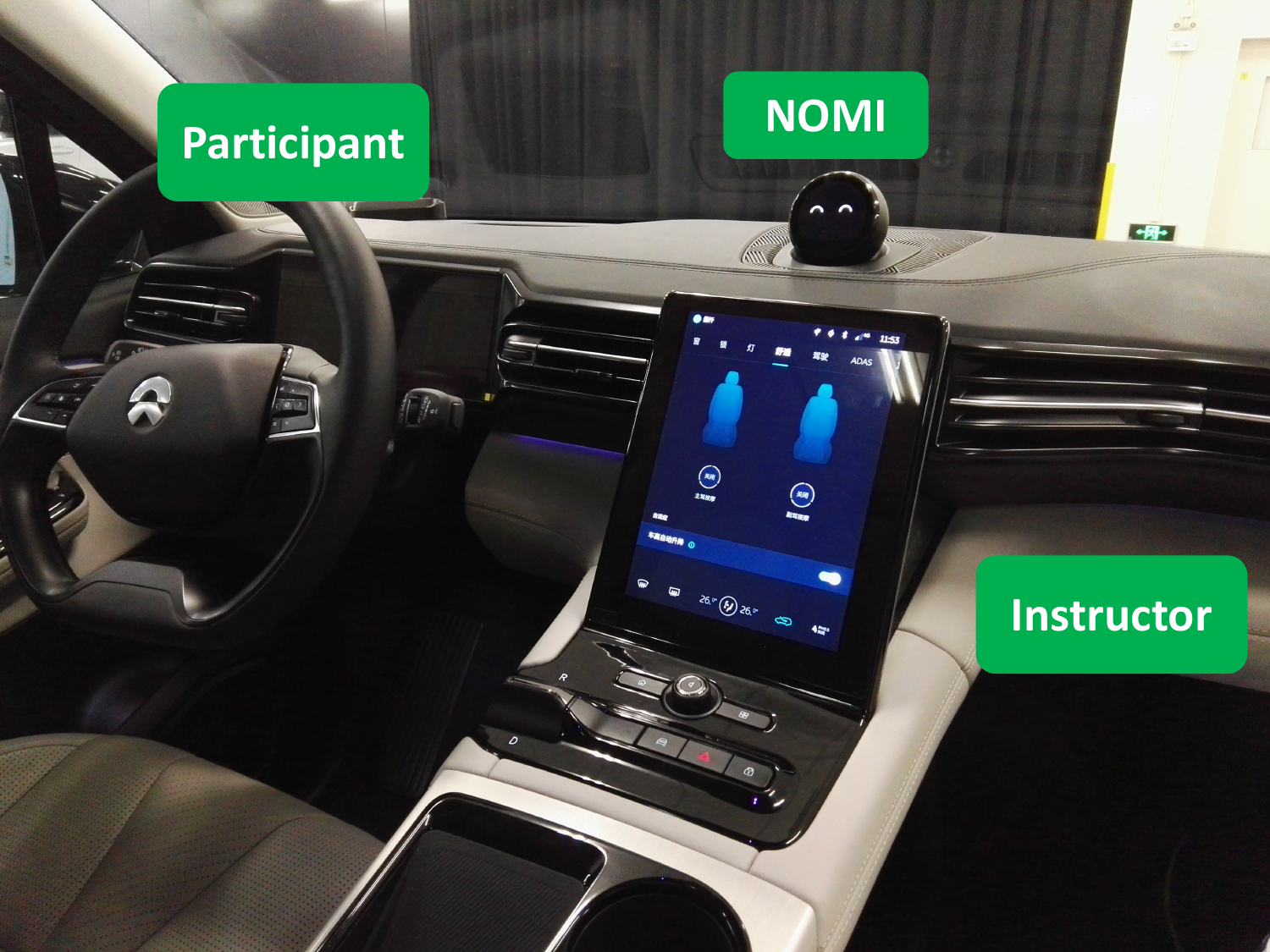}
  \includegraphics[scale=0.29]{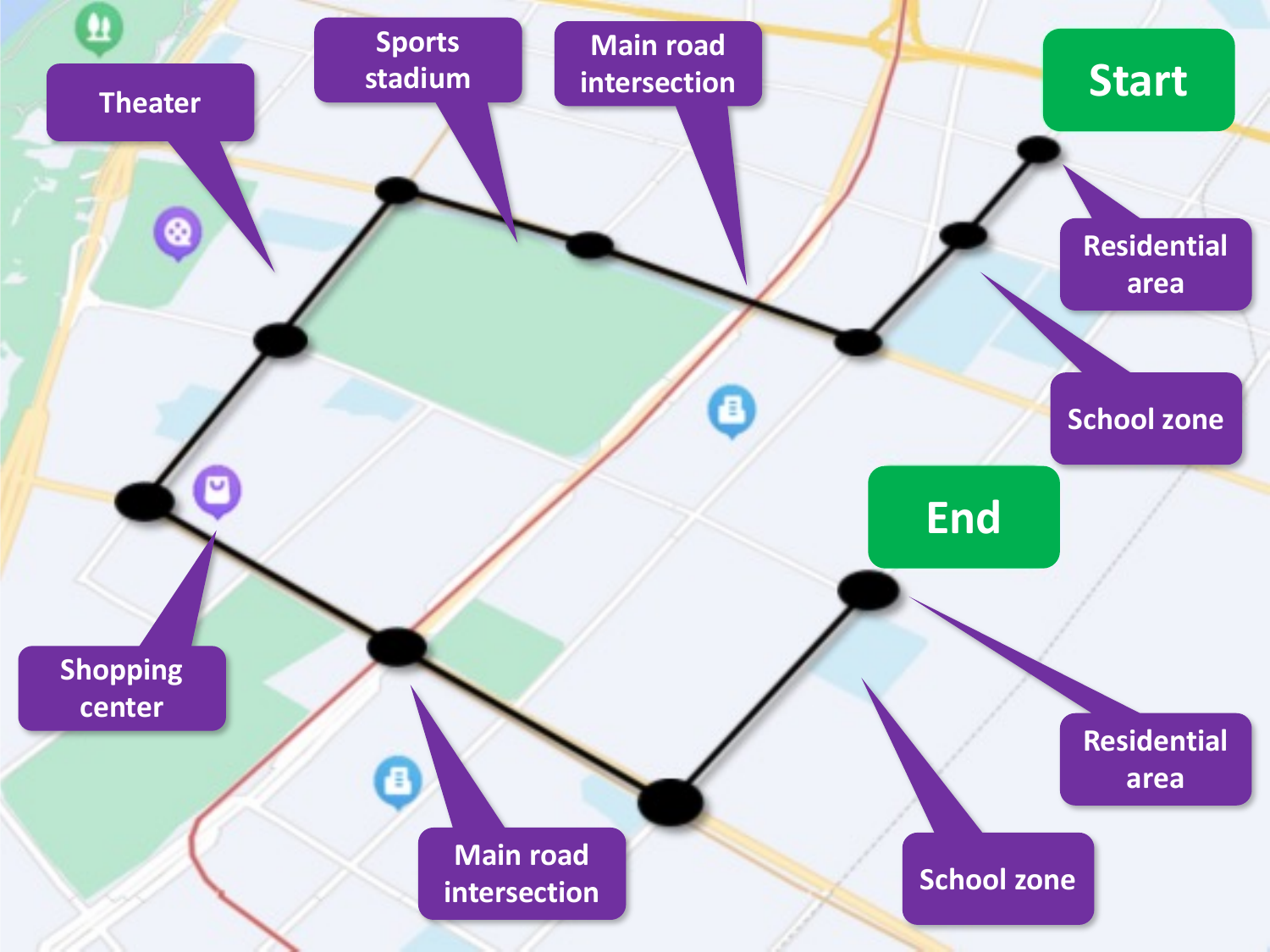}
  \caption{Environment of the non-driving experiments (left) and route of the on-road experiments (right).}
  \label{fig:experiment}
\end{figure*}

As the left in Fig.~\ref{fig:experiment}, the car was parked in a quiet parking lot and the participants issued voice commands while sitting in the driver's seat in the non-driving experiment. An instructor was sitting in the passenger seat guiding the tasks and recording videos for the evaluation analysis (e.g., calculating response time). In the on-road experiment, no instructor was sitting beside to avoid interfering with the effect of NOMI on the driving experience. The participants drive following the route as the right in Fig.~\ref{fig:experiment}. This route begins in a residential area with relatively low traffic. It then proceeds through a school zone, where reduced speed is required. Continuing further, it goes through densely populated areas, passing by a main road intersection, a sports stadium, a theater, a shopping center, and other crowded regions. Finally, it passes through another main road intersection, another school zone with reduced speed, and returns to a residential area for parking, concluding the journey. We intentionally selected this route to create diverse traffic conditions for a more comprehensive evaluation. The average time for each participant in the non-driving experiment is about 22 minutes, and for the on-road experiment, about 27 minutes.

The experiments employ both quantitative and qualitative measures to assess participants' experiences with NOMI. The quantitative assessment involves a questionnaire where participants rate user experience metrics on a five-point scale, from one (the lowest) to five (the highest). The qualitative measure resembles the semi-structured interview in the user study (Sec.~\ref{sec:interview}), involving participants answering prepared open-ended questions. Additionally, objective indicators are computed to evaluate NOMI's performance in the non-driving experiment, which serve as supplementary information to provide a comprehensive analysis of NOMI's effectiveness. Detailed information will be presented in the following sections. Note that we use the term `VA' when we particularly focus on the voice assistant functions (especially there is no NOMI presence). When referring to the VA and robot together, we use the term `RA'. Note that facial expressions cannot be pre-scripted or controlled, meaning we cannot pre-determine which facial expressions would accompany which task. However, it is precisely because of this random pattern that the effectiveness of NOMI can be fairly measured and demonstrated.

\subsection{Non-Driving Experiment}
In the non-driving experiment, all ten experimental tasks are mandatory. Notably, during task 3 \textit{navigation}, participants exclusively issue voice commands, with only the RA functioning while the car remains stationary. Similar to the user study, participants are informed before recording, retaining the right to request data deletion after analysis.

Furthermore, we explore the question: \textit{Can the robot NOMI visually influence participants' auditory perception of the in-car assistant?} by conducting a comparative experiment where NOMI is covered with a blanket, removing its visual presence. Additionally, we test the assistant's performance using both its cloud model (online mode) and local model (offline mode) to investigate its functionality in areas lacking stable network connectivity. Hence, the non-driving experiment encompasses four settings: online with and without NOMI presence, as well as offline with and without NOMI presence. The experiments are conducted in the order of \textit{offline $+$ w/o NOMI presence $\rightarrow$ offline $+$ w/ NOMI presence $\rightarrow$ online $+$ w/o NOMI presence $\rightarrow$ online $+$ w/ NOMI presence}. Participants are tasked with rating user experience across the following seven metrics.

\textbf{\textit{Recognition accuracy}} and \textbf{\textit{response accuracy}} measure a participant's satisfaction with the accuracy of the VA's speech recognition function. \textbf{\textit{Response time (voice)}} and \textbf{\textit{response time (action)}} measure a participant's subjective feeling of the delay (i.e., how long of the delay perceived) in the assistant's response. These four metrics are widely used in the subjective evaluation of VAs \cite{mitra2021analysis,cuadra2021my}. Besides, \textbf{\textit{mental workload}} and \textbf{\textit{auditory workload}} refer to the amount of cognitive resources and mental effort that a participant must allocate to complete the tasks, and the latter specifically involves auditory information processing (e.g., listen and comprehend) during the tasks. These two metrics are usually adopted in measuring driving experience \cite{de1996measurement,paxion2014mental,pauzie2008method}. \textbf{\textit{Engagement level}} means how actively and attentively a participant is involved or focused on the interaction with the assistant. It has been proven useful to measure the degree to which the participant is immersed or interested in Human-Computer Interaction (HCI) \cite{li2019expressing,inoue2021engagement}.

\subsubsection{Objective Indicators}
Before proceeding with the quantitative and qualitative measures of subjective ratings, the initial step involves calculating the \textit{\textbf{Word Error Rates (WERs)}} of the speech recognition results and assessing the \textit{\textbf{response time}} via video analysis. WER is a widely used metric for assessing the quality of speech recognition systems by comparing recognized words with the reference (i.e., ground truth) transcript \cite{wang2003word}, and is particularly important in applications such as transcription services and voice assistants \cite{li2023asr}. Typically expressed as a percentage, WER is calculated as:

\begin{align}
    WER = \frac{S + D + I}{N}
\end{align}

where $S$ represents the number of substitution errors (words that are incorrectly recognized). $D$ represents the number of deletion errors (words that are in the reference but missing in the recognized output). $I$ represents the number of insertion errors (extra words in the recognized output that are not in the reference). $N$ is the total number of words in the reference transcript. Lower WER values indicate better speech recognition performance.

On the other hand, response time in the context of VA refers to the time it takes for the system to provide a spoken response in the display after a user has issued a voice command or made a request. It measures the system's efficiency in processing and understanding the user's input and generating a relevant response \cite{shneiderman1984response}. Response time can vary based on two major factors: 1) system processing time, which involves speech recognition, natural language understanding, and intent recognition that process and interpret the user's spoken command or request; 2) action execution time, such as retrieving information from databases, accessing the internet, or controlling smart devices, that are for understanding the user's request \cite{stupak2009time,dabrowski201140}. Besides, network latency, processing power, and complexity of the request also have an impact on response time \cite{meyer1996duration}. A short response time is typically desirable in VA systems to provide a seamless and efficient user experience, and users generally expect quick and accurate responses to their voice queries. In this study, we measure the response time of both voice response and display change by the VA. Note that, some tasks do not respond with an action. For example, the display did not change to music play and the VA did not make a phone call when in offline mode. For these tasks, the end of speech recognition shown in the display is seen as the action response.

\begin{table*}[ht!]
\caption{WERs and response times of the ten experiment tasks. $\downarrow$: the lower the better.}
\begin{tabular}{lrrrrrr}
\toprule
\textbf{Experiment Task} & \multicolumn{2}{c}{\textbf{WER $\downarrow$}} & \multicolumn{2}{c}{\textbf{Response Time (voice) $\downarrow$}} & \multicolumn{2}{c}{\textbf{Response Time (action) $\downarrow$}} \\ 
 & Online & Offline & Online & Offline & Online & Offline \\ \midrule
\textit{Wake-up word} & 0.0\% & 0.0\% & 1.22s & 1.29s (\textit{+0.07s}) & 1.22s & 1.29s (\textit{+0.07s}) \\ 
\textit{POI search} & 1.8\% & 47.6\% & 2.42s & 5.15s (\textit{+2.73s}) & 2.41s & 5.16s (\textit{+2.75s}) \\ 
\textit{Navigation} & 0.0\% & 7.8\% & 2.31s & 3.41s (\textit{+1.10s}) & 2.30s & 3.42s (\textit{+1.12s}) \\ 
\textit{AC control} & 0.0\% & 25.8\% & 1.93s & 1.99s (\textit{+0.06s}) & 1.91s & 2.00s (\textit{+0.09s}) \\ 
\textit{Touchscreen function control} & 0.0\% & 32.2\% & 2.20s & 2.27s (\textit{+0.07s}) & 2.20s & 2.27s (\textit{+0.07s}) \\ 
\textit{Seat control} & 0.0\% & 30.7\% & 2.23s & 2.30s (\textit{+0.07s}) & 2.23s & 2.30s (\textit{+0.07s}) \\ 
\textit{Window, sunroof, sunshade control} & 0.0\% & 33.3\% & 2.37s & 2.42s (\textit{+0.05s}) & 2.34s & 2.43s (\textit{+0.09s}) \\
\textit{Dialing} & 0.0\% & 39.3\% & 2.54s & 2.67s (\textit{+0.13s}) & 2.54s & 2.69s (\textit{+0.15s}) \\ 
\textit{Entertainment} & 0.0\% & 28.9\% & 2.48s & 4.86s (\textit{+2.38s}) & 2.42s & 4.84s (\textit{+2.42s}) \\ 
\textit{Information inquiry} & 0.0\% & 37.5\% & 2.42s & 5.23s (\textit{+2.81s}) & 2.41s & 5.24s (\textit{+2.83s}) \\ \bottomrule
\end{tabular}
\label{tab:wer}
\end{table*}

Since all the participants are required to speak Mandarin, we hardly notice differences in their WERs and response times. We present the average WER and response time of all the ten participants in Table~\ref{tab:wer}. It can be observed that: \textbf{1)} There is hardly any discrepancy between the voice and action response times. This consistency ensures the validity of our subsequent analysis of the subjective ratings, as it will not be affected by the delay between the two responses. \textbf{2)} The online mode outperforms offline, which is natural as the cloud server has a much more powerful capability in processing the voice commands. \textbf{3)} The performances can vary significantly across the tasks. The WER and response time of \textit{wake-up word} consistently exhibit the best results, regardless of whether the VA is in online or offline mode. This is reasonable as all of the voice commands start from waking the VA up. It is crucial to ensure the speech recognition of the wake-up word `Hi, NOMI' is the most robust. Also, as this voice command is short and has no textual variability, a well-trained local model is sufficient, which has little difference from the cloud model. \textbf{4)} \textit{POI search} shows the worst WER, even in online mode. From the recorded video, we noticed that a shopping center name was incorrectly recognized, which is likely because it is an Out-of-Vocabulary (OOV) word to the speech recognition system. \textbf{5)} The performance of functions related to car control, such as \textit{AC control}, remains relatively stable regardless of the online and offline modes. This stability is attributed to the nature of these voice commands, which do not contain external information (e.g., locations, music) that necessitates server processing. On the contrary, other functions require more processing time, leading to a greater response delay.

\subsubsection{Quantitative Measure of Subjective Ratings}
Subsequently, we present the quantitative measures of the subjective ratings. We omit the individual ratings but show a summary of the results of the 2 (system mode: Online versus Offline) x 2 (NOMI presence: with NOMI versus without NOMI) between-subjects ANOVA in Table \ref{tab:anova}. These results are statistically significant, suggesting consistency among the participants. This consistency likely stems from the similarity in task content and environmental conditions, alongside nearly uniform responses from the RA (both VA and robot). Additionally, all participants possess previous experience with VAs, leading to shared knowledge regarding the functionality. Therefore, their ratings display minimal variation. 

The ANOVA results for the recognition and response accuracy indicate a significant main effect of the system being online or offline, but with no significant main effect of presence of NOMI, and no significant interaction between system status and robot presence. This suggests that the primary factor influencing outcomes is whether the system is online or offline, rather than the use of NOMI.

Regarding the results of response time for voice, the ANOVA results highlight significant main effects of system's being online or offline and the presence of NOMI, with no significant interaction effect. The post hoc analysis show that there are trends suggesting slight differences in response times between with and without NOMI (on-w/o NOMI vs on-w/ NOMI: \emph{p}-value=.052, off-w/o NOMI vs off-w/ NOMI: \emph{p}-value=.081), though these do not reach conventional levels of statistical significance. The significant differences between online and offline conditions, both with and without NOMI (on-w/o NOMI vs off-w/o NOMI: \emph{p}-value<.001, on-w/ NOMI vs off-w/ NOMI: \emph{p}-value<.001), emphasize the impact of system modes on response times. This effect is pronounced, indicating that the system being online significantly reduces the responses time for voice. 	

The ANOVA results also highlight significant main effects of system mode and the presence of NOMI on response times for action, without a significant interaction effect. The post hoc analysis further show that similarly to the response time for voice, the trends suggesting slight differences in response times between with and without NOMI (on-w/o NOMI vs on-w/ NOMI: \emph{p}-value=.052, off-w/o NOMI vs off-w/ NOMI: \emph{p}-value=.081), with a significant difference observed only when comparing online and offline conditions with NOMI (on-w/o NOMI vs off-w/o NOMI: \emph{p}-value<.193, on-w/ NOMI vs off-w/ NOMI: \emph{p}-value<.001).	

Moreover, the results highlight a significant interaction effect, evidenced by the post hoc analysis, indicates that the influence of NOMI on reducing mental workload is context-dependent, with its effect being notably different when comparing online to offline environments (on-w/o NOMI vs on-w/ NOMI: \emph{p}-value<.001, off-w/o NOMI vs off-w/ NOMI: \emph{p}-value<.001). The increase in mental workload from online to offline conditions, regardless of NOMI's presence, suggests that the offline context inherently demands more mental effort from users (on-w/o NOMI vs off-w/o NOMI: \emph{p}-value<.001, on-w/ NOMI vs off-w/ NOMI: \emph{p}-value<.001). 																																								
Similarly, the significant interaction effect on auditory workload in the ANOVA show that the influence of presence of NOMI depends on the system's mode. The post hoc analysis further show that the presence of NOMI significantly reduces auditory workload, especially in offline conditions (off-w/o NOMI vs off-w/ NOMI: \emph{p}-value<.001). Transitioning from online to offline conditions increases auditory workload, particularly when NOMI is not present (on-w/o NOMI vs off-w/o NOMI: \emph{p}-value<.001). 

Also, the results highlight a significant interaction effect on engagement levels. The post hoc analysis show that the presence of NOMI significantly increases engagement levels, both online and offline (on-w/o NOMI vs on-w/ NOMI: \emph{p}-value<.001, off-w/o NOMI vs off-w/ NOMI: \emph{p}-value<.001), with stronger effects observed in the offline condition. The transition from online to offline conditions significantly reduces the engagement level without NOMI's presence (on-w/o NOMI vs off-w/o NOMI: \emph{p}-value<.001). 

\begin{table}[ht]
\centering
\caption{Summary of two-way repeated measures ANOVA results.}
\label{tab:anova}
\begin{tabular}{@{}lcccc@{}}
\toprule
\textbf{Metrics} & \textbf{Measure} & \textbf{Online-Offline} & \textbf{w/-w/o NOMI} & \textbf{On-Off $*$ w/-w/o} \\
\midrule
\multirow{4}{*}{Recognition accuracy} & df & (1,9) & (1,9) & (1,9) \\
 & F & 418.846 & 1.0 & 1.0 \\
 & \emph{p}-value & \textbf{\textless.001} & .343 & .343 \\
 & $n^2$ & .979 & .1 & .1 \\
\hline
\multirow{4}{*}{Response accuracy} & df & (1,9) & (1,9) & (1,9) \\
 & F & 312.111 & 1.0 & 1.0 \\
 & \emph{p}-value & \textbf{\textless.001} & .343 & .343 \\
 & $n^2$ & .972 & .1 & .1 \\
\hline
\multirow{4}{*}{Response time (voice)} & df & (1,9) & (1,9) & (1,9) \\
 & F & 265.091 & 7.579 & 0.643 \\
 & \emph{p}-value & \textbf{\textless.001} & \textbf{.022} & .443 \\
 & $n^2$ & .967 & .457 & .067 \\
\hline
\multirow{4}{*}{Response time (action)} & df & (1,9) & (1,9) & (1,9) \\
 & F & 7.579 & 7.579 & 0.643 \\
 & \emph{p}-value & \textbf{.022} & \textbf{.022} & .443 \\
 & $n^2$ & .457 & .457 & .067 \\
\hline
\multirow{4}{*}{Mental workload} & df & (1,9) & (1,9) & (1,9) \\
 & F & 256.0 & 253.5 & 7.579 \\
 & \emph{p}-value & \textbf{\textless.001} & \textbf{\textless.001} & \textbf{.022} \\
 & $n^2$ & .966 & .966 & .457 \\
\hline
\multirow{4}{*}{Auditory workload} & df & (1,9) & (1,9) & (1,9) \\
 & F & 24.934 & 72.429 & 7.23 \\
 & \emph{p}-value & \textbf{.001} & \textbf{\textless.001} & \textbf{.025} \\
 & $n^2$ & 0.735 & 0.889 & 0.445 \\
\hline
\multirow{4}{*}{Engagement level} & df & (1,9) & (1,9) & (1,9) \\
 & F & 36.0 & 361.0 & 7.364 \\
 & \emph{p}-value & \textbf{\textless.001} & \textbf{\textless.001} & \textbf{.024} \\
 & $n^2$ & .8 & .976 & .45 \\
\bottomrule
\end{tabular}
\end{table}

\begin{figure*}[ht]
  \centering
  \includegraphics[width=0.995\textwidth]{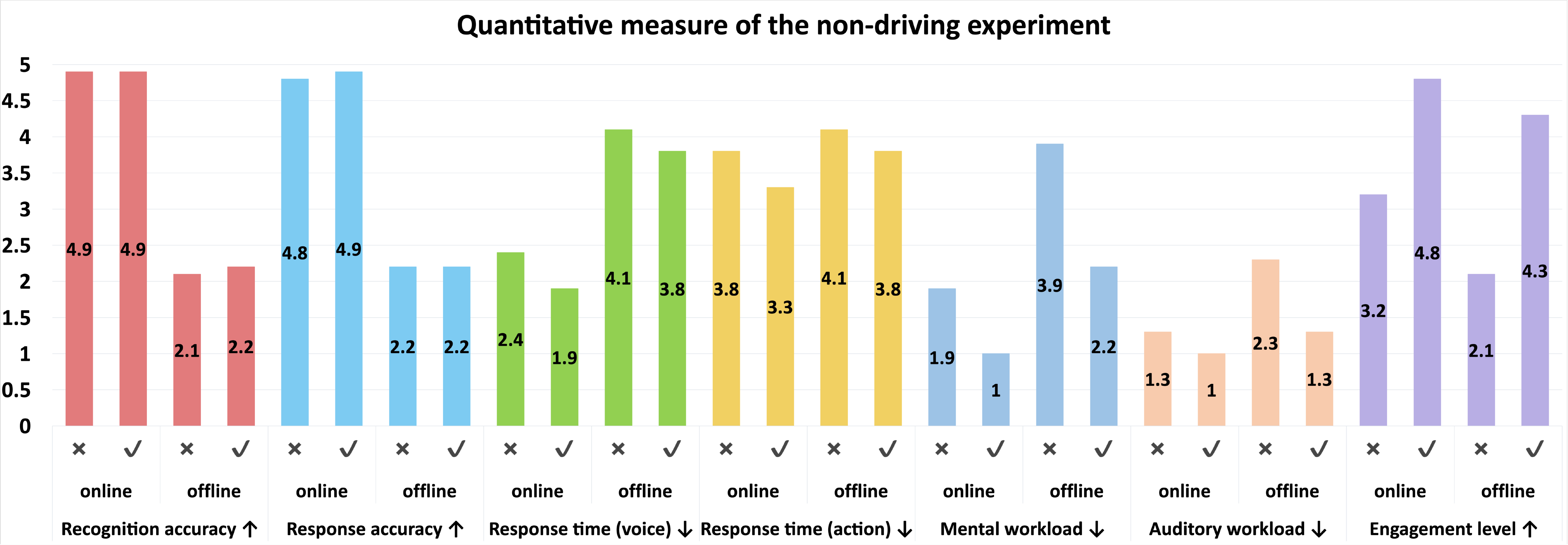}
  \caption{Subjective ratings of the seven metrics in online and offline modes. $\checkmark$ and $\times$ denote with and without NOMI presence, respectively. $\uparrow$: the higher the better. $\downarrow$: the lower the better.}
  \label{fig:non-driving}
\end{figure*}

The average scores of these subjective ratings are visualized in a bar chart graphic in Fig.~\ref{fig:non-driving}.
From both ~\ref{fig:non-driving} and \ref{tab:anova}, several observations can be made: \textbf{1)} In comparison to the offline mode, scores in every dimension are higher in the online mode, particularly in \textit{recognition accuracy}, \textit{response accuracy}, and \textit{response time}. This trend aligns with the patterns observed in the WERs and response times in Table~\ref{tab:wer}, indicating that a stable network connection significantly contributes to an improved subjective experience. \textbf{2)} When participants have visual access to NOMI, there are notable improvements in the ratings for \textit{mental workload}, \textit{auditory workload}, and \textit{engagement level}. This phenomenon is consistent with a previous study comparing robots versus voice assistants: Pollmann \emph{et al.} \cite{pollmann2020robot} provided indications that the perceived entertainment experience of a game can be increased when the game is played with a Pepper robot rather than with a voice assistant. However, their ratings concerning the VA's voice performance-related metrics remain largely unaffected. This finding suggests that, despite no change in the VA's voice performance, participants experience reduced mental and auditory load and increased engagement in the interaction when they have visual access to NOMI. This highlights NOMI's effectiveness in enhancing the driving experience.

\subsubsection{Qualitative Measure of Subjective Ratings}
\label{sec:qua-non}
Finally, we conduct a semi-structured interview with the ten participants, which consist of three parts of questions. Part I and III center on the RA (including both VA and NOMI), while Part II focuses on NOMI. The interview questions are detailed in Table~\ref{tab:non-driving}. Same as Section~\ref{sec:interview}, we present overall conclusions, highlighting selected representative or exceptional answers as examples.

\begin{table*}[ht]
  \caption{Interview questions for the non-driving experiment.}
  \label{tab:non-driving}
  \begin{center}
  \begin{tabular}{l}
    \toprule
    \textit{\textbf{Part I}} \\
    \hdashline
    Q1. Does the RA meet your expectations? \\ \ \ \ \ \ \ \ If yes, in what ways? If not, what are the reasons? \\
    Q2. Is the RA different from the VAs you have previously used? \\ \ \ \ \ \ \ \ If yes, in what ways? \\
    \midrule
    \textit{\textbf{Part II}} \\
    \hdashline
    Q3. Do you like NOMI? \\ \ \ \ \ \ \ \ If yes, what aspects do you like it? If not, what are the reasons? \\
    Q4. Do you agree that NOMI improves your user experience? \\ \ \ \ \ \ \ \ If yes, in what ways? If not, what are the reasons? \\
    \midrule
    \textit{\textbf{Part III}} \\
    \hdashline
    Q5. What additional features do you think the assistant should have?  (in terms of both voice and robot)  \\ \ \ \ \ \ \ \ What are the reasons? \\
    \bottomrule
  \end{tabular}
  \end{center}
\end{table*}

\textbf{Part I - Q1}.
It is impressive that all participants express satisfaction with the RA, with many stating it exceeds their expectations, largely due to NOMI's presence. Notably, participants like P2 have shifted their impressions significantly due to this setup. However, participants also highlighted a discrepancy between the online and offline modes. This suggests a notable impact on their perception based on the system's connectivity status.

\textit{``For sure. It's indeed amazing. I feel like the car has become more than just a means of transportation. It's essentially another interactive platform with the presence of NOMI. This is a completely different experience from other VAs that I have ever used. However, when there was no network connection, the VA was dumb.'' (P2)}

\textit{``The VA is very advanced. I really didn't expect it. It can be said to far exceed my expectations. I've never used a car-based VA before. I've always used my smartphone. I didn't expect the conversation with the in-car VA to be so smooth. However, I don't think it can be used in offline mode.'' (P13)}

\textit{``The robot is so fun. Can I buy it for home use? I sometimes drive as a DiDi (i.e., Chinese Uber) driver. With the robot, it must be very fun. I think the passengers would like it as well.'' (P24)}

It is evident that participants are impressed by the assistant's interactivity in the online mode and hold a strong appreciation for NOMI. However, there is a clear contrast when the assistant operates without a network connection yet NOMI is expected to alleviate the uncomfortableness according to the subjective ratings.

\textbf{Part I - Q2}.
It appears that responses vary based on participants' prior experience with smart home VAs. Those accustomed to using VAs for controlling devices like curtains and lighting do not find the in-car VA particularly impressive. Conversely, participants without smart home VA experience view the in-car VA as somewhat unique. We are considering conducting a correlation analysis between the familiarity/accustomization with VAs and the degree to which they find the RA impressive for future work. Nevertheless, they all acknowledge the advantage of having an in-car VA with a robot appearance.

\textit{``I feel it's quite similar to the Xiao AI I have at home. I use a set of Xiaomi products, many of which can be controlled using voice commands. But I am still willing to talk to it. The robot makes a difference.'' (P5)}

\textit{``It does feel somewhat different from what I've used before. This one seems to be more focused on car-related functions, which are not supported by CarPlay. Besides, at home, I typically use the VA for checking the weather, playing music, and having casual conversations. On my phone with Siri, I usually use it for asking questions or looking up information. However, with this in-car VA, the emphasis seems to be on navigation and in-car entertainment. I don't really think about using it to search for information. However, having a conversation with NOMI seems fun. It's probably due to the novelty, but I'm quite eager to see how it responds to my commands, and I'm curious about how its facial expressions might change.'' (P7)}

\textit{``My car also has a VA but does not support as many functions as this one. I feel like every necessary part can be controlled via voice.'' (P13)}

These responses suggest that users with different VA experiences hold opposite opinions regarding functionality. Some believe this in-car VA does not significantly differ from other VAs they use at home. However, due to its tailored design for in-car use, some participants still feel it is advanced. Features like \textit{POI search} and \textit{navigation} hold particular significance, marking them as key areas for future development and enhancement. Additionally, NOMI continues to play a role in encouraging engagement with the VA.

\textbf{Part II - Q3}.
It is anticipated that all participants favored NOMI, aligning with the subjective ratings in Fig.~\ref{fig:non-driving}.

\textit{``It's really nice! The appearance, movement, and facial expression are so cute.'' (P6)}

\textit{``It is special. It's like a little cat. Even when you're not talking to it, its face still moves on its own.'' (P14)}

\textit{``Yes, I like it! This is quite fun! How come other in-car assistants don't have this kind of robot?'' (P23)}

Although we have expected this result, it is noteworthy that NOMI greatly attracts participants' attention while using the RA. This finding emphasizes the crucial role of integrating a social robot in the development of in-car VAs.

\textbf{Part II - Q4}.
The responses to this question exhibit remarkable consistency. Every participant unanimously agrees that NOMI significantly enhances their in-car assistant experiences. They all note that with NOMI, their interactions with the assistant became more engaging and interactive.

\textit{``Yes, definitely! I was more willing to talk to the VA with NOMI and more patient to wait for its response.'' (P2)}

\textit{``I agree. NOMI is incredibly lifelike. You know it's mechanical and robotic, but it still gives you a sense of being alive.'' (P5)}

\textit{``Of course yes. As said, NOMI is like a cat. Talking to it feels like talking to a pet.'' (P14)}

\textit{``With NOMI, I can visualize what the VA looks like. And once I have a visualization, I start to develop an attachment to it.'' (P24)}

Clearly, NOMI's presence significantly enhances the user experience. It not only gives the VA a tangible form but also sparks a desire for interaction, elevating the VA to a higher stage—RA. Some users even perceive certain pet-like characteristics in the interaction.

\textbf{Part III - Q5}.
Certain features are consistently highlighted by multiple participants, notably emotion interaction, a concept more widely comprehended by the general public. Moreover, participants with a certain level of IT knowledge offer suggestions for a few specific features.

\textit{``It would be more impressive if NOMI could change its emotion based on mine, like a human. Most of the time, NOMI expresses a happy face and voice. A variable expression can make it more human-like.'' (P2)}

\textit{``It still gives me a somewhat robotic impression. For example, the style of its responses is quite uniform, and its tone of speech also lacks variation. Also, it would be even better if it could remember previous conversations for an extended period to facilitate multi-turn dialogue understanding.'' (P7)}

\textit{``It seems that no matter how I speak, NOMI always uses a cheerful tone. It would be more interesting if it could give warnings in a stern tone maybe when I'm driving at a higher speed.'' (P24)}

We can see that participants still have some expectations for NOMI. One of the key areas for future improvement in NOMI is enhancing its human-like interaction capabilities. This aspect has also been addressed in previous HCI research, including understanding disfluencies (e.g., fillers) \cite{li2022robotic}, adapting emotions based on the user's \cite{li2023know}, and fostering shared laughter with the user \cite{inoue2022can}. These aspects play a vital role in establishing rapport in HCI.

Given that the non-driving experiment primarily focuses on VA usage without any driving aspects, we conduct an on-road experiment to investigate whether the social robot could similarly have a substantial impact on the user experience during driving.

\subsection{On-Road Experiment}

\subsubsection{Quantitative Measure of Subjective Ratings}
In the on-road experiment, participants are allowed to issue voice commands as they wish, simulating their typical driving scenarios. However, the first three tasks outlined in Section~\ref{sec:task} are mandatory--\textit{wake-up word}, \textit{POI search}, and \textit{navigation}. To conduct a comparative analysis, we cover NOMI with a blanket and instruct the participants to drive without seeing NOMI for the first half of the route. In the second half, participants are required to remove the blanket and continue driving. As previously mentioned in Section~\ref{sec:task}, we intentionally select this route, partly because the driving environments in the first and second halves are nearly identical, encompassing a residential area, a school zone, and a long crowded road. This choice is made to ensure a relatively fair comparison for driving with and without NOMI. Given that the impact of network connection is not the primary focus of this experiment, and asking participants to perform two driving sessions (both online and offline modes) would impose an undue burden, which may affect their driving experience, we exclude the online-offline comparison in the on-road experiment with the network connected all the time.

The participants need to rate the user experience in the following eleven metrics on a scale of five. In the same manner as the non-driving experiment, a rating of one indicates the lowest, while five indicates the highest.

\textbf{\textit{Trust}} was initially introduced to assess the driver's trust in an automated driving system \cite{hergeth2016keep}. In this study, we use this metric to measure the extent to which a participant trusts the assistant responses. \textbf{\textit{Attractiveness}} \cite{laugwitz2008construction} is utilized to assess a participant's views on the appearance and design of NOMI, as well as the voice of the assistant. \textbf{\textit{Stimulation}} \cite{laugwitz2008construction} is adopted to measure whether a participant experiences excitement or stimulation from the assistant. \textbf{\textit{Usefulness}} and \textbf{\textit{satisfaction}} \cite{van1997simple} are used to assess the level of contentment or approval from a participant after using the assistant, and whether a participant considers the assistant to have practical value and utility for the tasks. \textbf{\textit{Interference}} and \textbf{\textit{stress}} \cite{pauzie2008method} evaluate the potential disturbances caused by the assistant, as well as the levels of constraints and stress factors such as fatigue, feelings of insecurity, irritation, discouragement, and so forth, experienced during the on-road experiment. Last, we also measure \textbf{\textit{mental workload}}, \textbf{\textit{auditory workload}}, and \textbf{\textit{engagement level}} as the non-driving experiment. These metrics have proven useful for measuring user experience in driving \cite{braun2019your}.

\begin{table}[ht]
\centering
\caption{Paired t-test results for impact of NOMI presence on user perceptions.}
\label{tab:ttest}
\begin{tabular}{lccccc}
\toprule
\textbf{Metrics} & \textbf{NOMI} & \textbf{Mean} & \textbf{SD} & \multicolumn{1}{c}{\textbf{\emph{t}(9)}} & \multicolumn{1}{c}{\textbf{\emph{p}-value}} \\
\midrule
\multirow{2}{*}{Trust} & $\times$ & 2.6 & 0.699 & {$-7.965$} & \textbf{{\textless.001}} \\
 & $\checkmark$ & 4.3 & 0.483 & & \\
\addlinespace
\multirow{2}{*}{Attractiveness} & $\times$ & 1.5 & 0.527 & {$-12.829$} & \textbf{{\textless.001}} \\
 & $\checkmark$ & 4.7 & 0.483 & & \\
\addlinespace
\multirow{2}{*}{Stimulation} & $\times$ & 1.6 & 0.516 & {$-15.000$} & \textbf{{\textless.001}} \\
 & $\checkmark$ & 4.1 & 0.316 & & \\
\addlinespace
\multirow{2}{*}{Usefulness} & $\times$ & 2.3 & 0.483 & {$-4.583$} & \textbf{{.010}} \\
 & $\checkmark$ & 3.7 & 0.674 & & \\
\addlinespace
\multirow{2}{*}{Satisfaction} & $\times$ & 2.4 & 0.516 & {$-8.820$} & \textbf{{\textless.001}} \\
 & $\checkmark$ & 4.6 & 0.516 & & \\
\addlinespace
\multirow{2}{*}{Interference} & $\times$ & 2.9 & 0.737 & {$2.449$} & \textbf{{.037}} \\
 & $\checkmark$ & 2.5 & 0.527 & & \\
\addlinespace
\multirow{2}{*}{Stress} & $\times$ & 2.1 & 0.567 & {$3.000$} & \textbf{{.015}} \\
 & $\checkmark$ & 1.6 & 0.516 & & \\
\addlinespace
\multirow{2}{*}{Mental workload} & $\times$ & 2.3 & 0.674 & {$2.333$} & \textbf{{.045}} \\
 & $\checkmark$ & 1.6 & 0.516 & & \\
\addlinespace
\multirow{2}{*}{Auditory workload} & $\times$ & 1.5 & 0.527 & {$1.964$} & {.081} \\
 & $\checkmark$ & 1.2 & 0.421 & & \\
\addlinespace
\multirow{2}{*}{Engagement} & $\times$ & 1.6 & 0.699 & {$-6.332$} & \textbf{{\textless.001}} \\
 & $\checkmark$ & 3.0 & 0.666 & & \\
\bottomrule
\end{tabular}
\end{table}

\begin{figure*}[ht]
  \centering
  \includegraphics[width=0.995\textwidth]{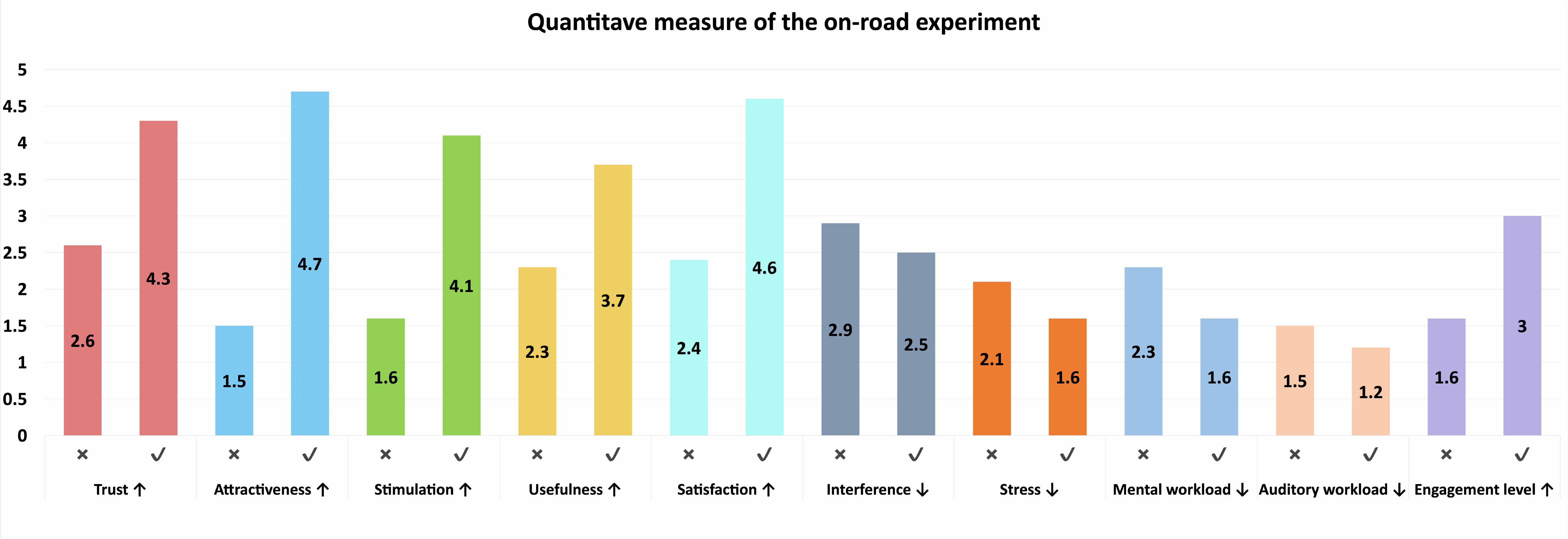}
  \caption{Subjective ratings of the eleven metrics. $\checkmark$ and $\times$ denote with and without NOMI presence, respectively. $\uparrow$: the higher the better. $\downarrow$: the lower the better.}
  \label{fig:on-road}
\end{figure*}

Since there in no online-offline comparison, we implement a paired t-test instead of the two-way repeated measures ANOVA. The results are shown in Table~\ref{tab:ttest}, which demonstrate the positive effects brought by NOMI. The paired t-test results revealed significant differences, suggesting that NOMI significantly enhanced all metrics except for auditory workload. The results are also visualized in Fig.~\ref{fig:on-road} for a clearer comparison. It is noticeable that having visual access to NOMI enhances every user experience metric, albeit to varying extents. Notably, improvements in \textit{interference}, \textit{stress}, \textit{mental workload}, and \textit{auditory workload} are not as substantial, especially as \textit{interference} and \textit{auditory workload} demonstrate less sensitivity to NOMI. This observation is plausible, as their scores are initially not particularly high without NOMI presence. Moreover, the small decrease in workloads remains consistent with the online mode in the non-driving experiment, indicating that NOMI might not significantly contribute to reducing workload when the VA works smoothly.

\subsubsection{Qualitative Measure of Subjective Ratings}
We perform follow-up semi-structured interviews comprising three parts. Part I and III contain questions about the RA (including both VA and NOMI), while Part II pertains to NOMI. The interview questions are detailed in Table~\ref{tab:on-road}. We outline the general conclusions and highlight selected representative or exceptional answers as examples. The majority of the responses are selected from the same participants in Section~\ref{sec:qua-non} to ensure a fair comparison and analysis.

\begin{table*}[ht]
  \caption{Interview questions for the on-road experiment.}
  \label{tab:on-road}
  \begin{center}
  \begin{tabular}{l}
    \toprule
    \textit{\textbf{Part I}} \\
    \hdashline
    Q1. Does the driving experience with the RA meet your expectations? \\ \ \ \ \ \ \ \ If yes, in what ways? If not, what are the reasons? \\
    Q2. Does the functionality of the RA differ between driving and stationary? \\
    \midrule
    \textit{\textbf{Part II}} \\
    \hdashline
    Q3. Do you believe NOMI enhances your driving experience? \\  \ \ \ \ \ \ \ If yes, in what ways? \\
    \midrule
    \textit{\textbf{Part III}} \\
    \hdashline
    Q4. What additional features do you think the assistant should have?  (in terms of both voice and robot) \\ \ \ \ \ \ \ \ What are the reasons? \\
    \bottomrule
  \end{tabular}
  \end{center}
\end{table*}

\textbf{Part I - Q1}.
Nine out of ten participants responded positively, while only one provided neutral feedback.

\textit{``For sure. It's so fun driving with the RA and talking to it. Especially when I was speaking, NOMI was looking at me. I felt accompanied.'' (P2)}

\textit{``A truly fantastic experience driving with it. It exceeded my expectations. As I mentioned before (in the non-driving experiment), I didn't expect the conversation to be so smooth. I never thought I would engage in conversation with a VA while driving.'' (P13)}

\textit{``Yes, it meets my expectations. I usually enjoy talking to passengers, and now with the robot, I can chat with it while driving alone.'' (P24)}

\textit{``It's alright. It meets my expectations just because I didn't set it very high. The VA is essentially similar to the one of MBUX, just with a robot appearance.'' (P7)}

From the participant feedback, we can observe that most participants are very satisfied with the RA, especially because of NOMI's presence. They are more willing to engage in conversations because NOMI provides a human-like sense of companionship. However, for the participants who have relevant knowledge and rich experience with VAs, the assistant (even with NOMI) seems not to be more attractive than previous VAs they have used. Since P7 differed from other participants in the previous interview study by consistently providing a somewhat neutral to negative evaluation, we believe P7 might not represent the mainstream opinion. However, P7's feedback indeed highlights certain shortcomings of the RA, which should be thoroughly considered as one of the directions for the future development of all in-car assistants.

\textbf{Part I - Q2}.
The participants generally provide consistent feedback regarding functionality. However, some note a distinct impact due to NOMI's presence. As our emphasis on NOMI lies in Part II, we intervene and redirect participants to Q3 whenever they begin discussing NOMI.

\textit{``It's convenient to use while driving, but I still don't feel it's different from the home-use VAs. Specifically, I don't like controlling the AC via voice commands. Using a button would be much faster for manipulation.'' (P5)}

\textit{``Well, the functionality doesn't have much difference. The major difference lies in NOMI. I saw it nodding and changing expressions from time to time, it's so lifelike. I think this really helps alleviate the loneliness of driving alone.'' (P7)}

\textbf{Part II - Q3}.
All participants provided positive responses, aligning with our expectations.

\textit{``Definitely! I felt more relaxed and cheerful. It looked at me when I was talking, and sometimes, when waiting at red lights, it made expressions as if humming and playing the guitar. So fun!'' (P2)}

\textit{``Yes. I don't drive much, so I get a bit anxious in congested areas. Usually, when my boyfriend is in the passenger seat, I feel more at ease. Just now, near the shopping mall, cars were queuing to enter the underground parking, and half of the road was occupied. I was afraid of scratching the neighboring cars. If I were alone, I wouldn't have the courage to drive, but NOMI made me feel relatively secure. It's like having someone accompanying me, which I find particularly comforting. I think the feeling of having a companion is very important, especially for females like me who feel scared driving alone, especially at night.'' (P6)}

\textit{``It's like having a pet or a child that won't bother you. If you want to talk to it, it keeps you company. If you don't seek it out, it quietly stays by your side. It's fun, adding a bit of enjoyment to driving.'' (P14)}

We can see that the participants mentioned `companionship' or expressed similar sentiments, indicating that NOMI can reduce their sense of loneliness, add enjoyment, and make the drivers feel more secure or less stressed. Previous research has demonstrated that introducing empathy to drivers can significantly reduce their stress \cite{hernandez2014autoemotive}, partially affirming the empathetic capability of NOMI. This is a crucial factor in the design of human-machine interaction products, highlighting NOMI's excellent performance in this aspect. We also have reason to believe that `companionship' is a crucial consideration in future car assistant designs.

\textbf{Part III - Q4}.
Although participants generally agree that NOMI enhances the driving experience and have already provided feedback in the non-driving experiment, some participants still highlight areas for improvement based on their on-road driving experience.

\textit{``I believe it would be better if it could engage in multi-turn and continuous conversations in certain situations. I have a 3-year-old child who always wants to talk to me when I'm driving, which can be a bit distracting. I think he would really enjoy talking to NOMI, but the current question-answer format doesn't satisfy a child's curiosity. If the assistant could ask some follow-up questions based on what the child says and if NOMI could show more interesting facial expressions, it might be more engaging.'' (P2)}

\textit{``I'm thinking that it could potentially adapt to different users' speaking habits to personalize its features. For those who enjoy listening to music while driving, the assistant could directly ask them what music they'd like to play once they get into the car, and then NOMI can somewhat shake its head with the rhythm. For people with less-than-ideal driving habits, the assistant can occasionally offer reminders with NOMI showing a red angry face.'' (P7)}

These comments align with current trends in HCI research. For instance, the exploration of users' customization and personalization preferences is a widely researched area across various user groups \cite{zargham2022want, yuan2019speech}. Additionally, the aspect of a multi-turn dialogue function suggested by P2 was previously mentioned by P7 in Section~\ref{sec:qua-non}, indicating potential common needs across different users.

\subsection{Summary}
The experimental evaluation provides several major findings: \textbf{1)} The user experience toward in-car VAs is greatly influenced by the network condition. In situations with good network connectivity, the user feedback is generally positive, whereas in poor network conditions, the user experience significantly decreases. However, in the presence of NOMI, this social robot can ameliorate this situation, especially in terms of users' subjective experience. Nevertheless, NOMI doesn't significantly enhance their experience with the objective performance of the VA, such as \textit{recognition accuracy}. \textbf{2)} The users' evaluations of the functionality of the R A show minimal differences between stationary and driving scenarios. However, the users' focus slightly differs in these two situations. When the car is stationary, users are more concerned about the robot's variety of facial expressions and tones. Yet, while driving, they prioritize the VA's dialogue capabilities. This seems reasonable as, when the car is stationary, the assistant resembles a regular home VA where conversations are initiated by users, and their attention is more on the sentence-level content of the VA's responses. However, while driving, users are more focused on the road and vehicle control, paying less attention to the lower-level aspects of speech (i.e., spoken content and prosody variation) but being more sensitive to the higher-level aspects (i.e., multi-turn and memory). \textbf{3)} In scenarios where the speech performance is identical, NOMI effectively enhances the users' driving experience, and many users attribute this enhancement to the `companionship' that NOMI provides. This finding suggests that users require an RA during driving that does not disturb them while offering a sense of companionship. \textbf{4)} Even certain users who did not show a preference for using VAs (P2) or have rich experience and background with VAs (P7) were either impressed by the superiority of the user experience with the in-car VA with NOMI or acknowledged the usefulness of NOMI's presence after using it. Based on these findings, we believe that a well-designed robot can largely enhance the in-car VA and we hope these findings can offer valuable insights for the future design of in-car assistants.

\section{Personality Assessment}
\label{sec:personality}

Considering the importance of personality evaluation in VAs \cite{braun2019your,poushneh2021humanizing,lopatovska2020personality,christensen2021your}, the ten participants are asked to rate the RA's perceived personality compared to other in-car VAs they have encountered, after the above experimental evaluation. Personality can be defined using Fiske’s Warmth-Competence model \cite{fiske2002model} and the Big Five personality traits, often referred to as OCEAN (openness, conscientiousness, extraversion, agreeableness, neuroticism) \cite{john1999big}. Past studies have typically evaluated one or the other, yet in this work, we adopt both models for a comprehensive analysis. Each personality dimension is rated on a scale of five (one denotes the lowest and five the highest).

\begin{figure*}[ht]
  \centering
  \includegraphics[scale=0.341]{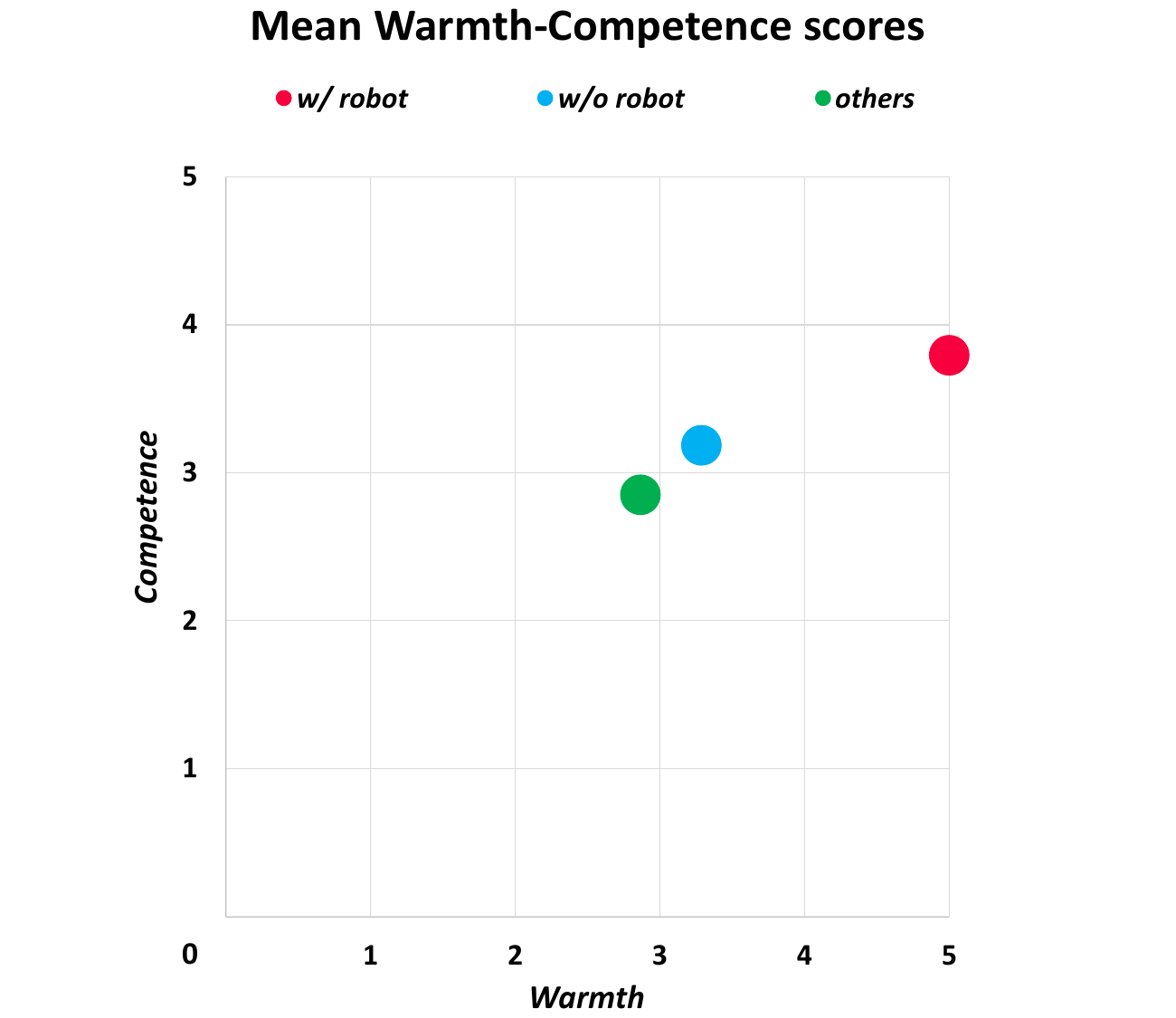}
  \includegraphics[scale=0.341]{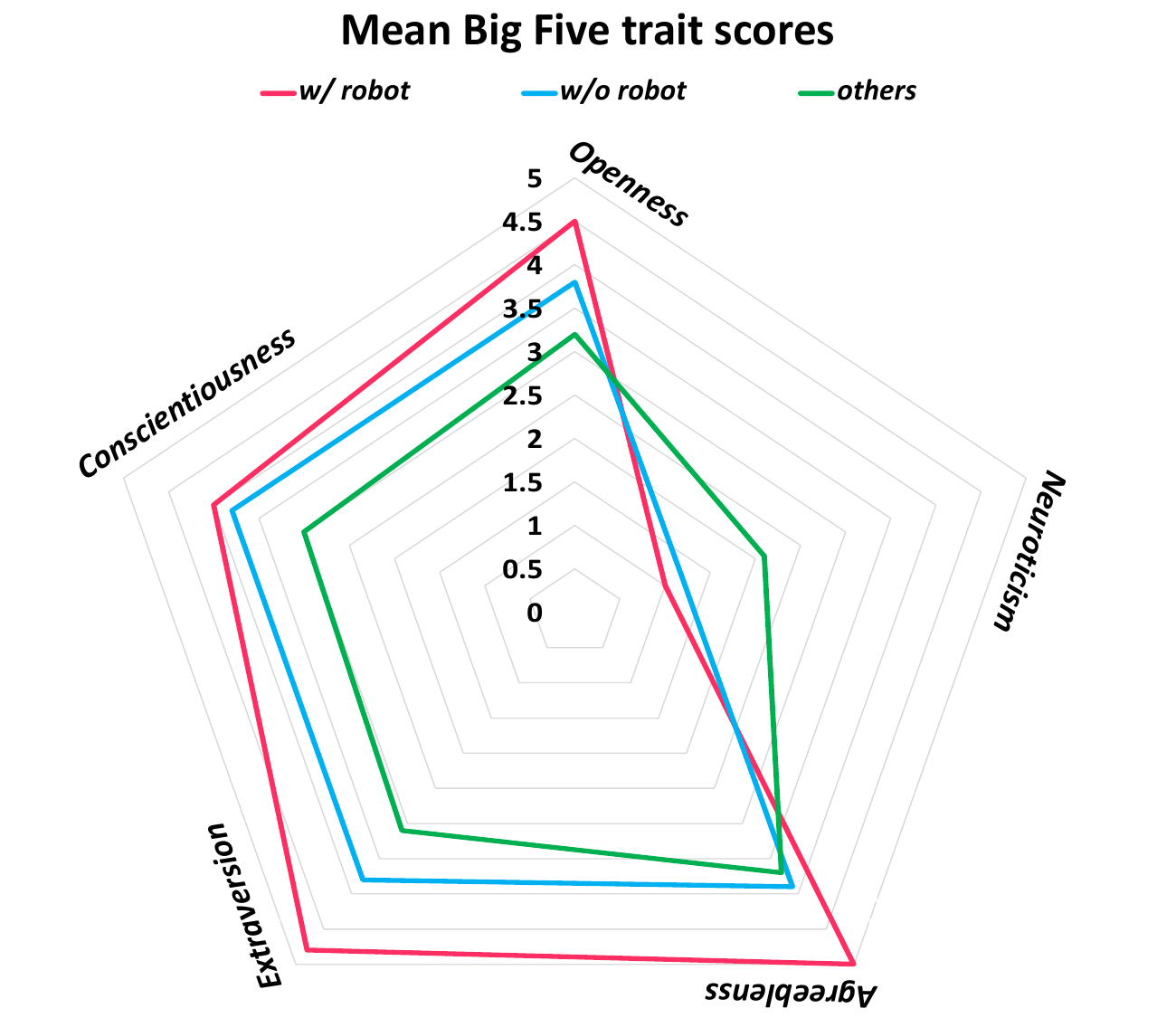}
  \caption{Mean scores of the personality dimensions. Warmth-Competence model (left) and Big Five trait (right).}
  \label{fig:personality}
\end{figure*}

Given that participants have varied experiences with different VAs, we request them to compare the RA to their most frequently used VA, collectively referred to as \textit{others}. Therefore, three types of assistants are used for comparison: 1) \textit{w/ robot NOMI}, 2) \textit{w/o robot NOMI}, and 3) \textit{others}. Table~\ref{tab:personality} shows the mean scores of each personality dimension and Fig.~\ref{fig:personality} visualizes them. It can be noted that \textbf{1)} Participants have a high impression of \textit{warmth}, \textit{extraversion}, and \textit{agreeableness} when interacting with NOMI. This aligns with previous research indicating that individuals are more likely to establish rapport with a human-like machine \cite{riek2010my,staffa2016recommender}. \textbf{2)} Conversely, when without NOMI, the scores for these three dimensions largely decrease, dropping by more than 1.0. However, certain dimensions do not exhibit a strong correlation with the robot's appearance. For instance, the scores for \textit{conscientiousness} and \textit{neuroticism} show minimal variation. This might be attributed to the significant influence of voice on these dimensions, given their high correlation with arousal which indicates high voice energy) \cite{o1996interactional}.
\textbf{3)} The assistant, when without NOMI, exhibits personality traits comparable to other VAs. This illustrates the challenge of expressing a robot's personality without a seamless alignment between its appearance and voice. This observation supports a viewpoint articulated in a study on robot identity \cite{li2022robotic}: giving a mismatched voice to a robot might introduce a confounding effect.

\begin{table*}[ht!]
\caption{Mean scores of personality dimensions.}
\label{tab:personality}
\begin{threeparttable}
\begin{tabular}{lrrrrrrr}
\toprule
\textbf{}               & \textbf{Warm.} & \textbf{Comp.} & \textbf{O} & \textbf{C} & \textbf{E} & \textbf{A} & \textbf{N} \\ \midrule
\textit{w/ robot}  & 5.0        & 3.8        & 4.5        & 4.0        & 4.8        & 5.0        & 1.1        \\
\textit{w/o robot} & 3.3        & 3.2        & 3.8        & 3.8        & 3.8        & 3.9        & 1.2        \\
\textit{others}         & 2.9        & 2.9        & 3.2        & 3.0        & 3.1        & 3.7        & 2.1        \\ \bottomrule
\end{tabular}
\begin{tablenotes}
\footnotesize
\item[$*$]All \textit{p} values < 0.05.
\end{tablenotes}
\end{threeparttable}
\end{table*}

\section{Risks of In-Car Robot Assistants}
\label{sec:concerns}

Risks such as privacy, security, and ethical concerns have been intensively investigated in VAs \cite{cheng2022personal}. However, the in-car scenario has only recently gained attention. To our knowledge, these issues have been examined primarily through empirical studies \cite{liu2023exploring} and driving simulators \cite{kim2021please}. Therefore, we aim to pioneer the study in real driving scenarios.

Initially, we pose an open-ended question to all 30 participants: \textit{Do you have any concerns about using an in-car assistant with a robot like NOMI?} For participants from the US, UK, and Japan who did not take part in the user experiments, we conduct interviews, providing them with information about NOMI and showing an official promotional video\footnote{https://www.youtube.com/watch?v=SAZ2Dd9lrVc}. We deliberately did not mention issues like ethics and privacy, leaving participants to contemplate on their own. Some participants were concerned about road safety issues arising from the distraction caused by NOMI, which is in line with a potential risk raised by a previous study \cite{lee2001speech}. However, only a few participants were aware of issues related to privacy, surveillance, and data security.

\textit{``Does this robot have a camera? I'm a bit hesitant about anything with a camera. I always cover the camera on my laptop. I saw in the YouTube video (the official promotional video) that there is a camera for taking pictures. If there is a camera, I might not be very willing to use it, or I would cover the camera.'' (P1)}

\textit{``Although VAs are convenient, to be honest, I've never been particularly trusting them. The things I say are all eavesdropped by the software on my phone. Some shopping apps (we omit the names) automatically push related products to me. I'm afraid the in-car assistants would do the same.'' (P2)}

\textit{``I am concerned about distracted attention which may cause traffic accidents while driving. Other than that, it seems fine to me.'' (P3)}

\textit{``If my voice data, driving data, and facial data, among other information, are all stored by a third party, I am uneasy about the storage of my data. However, this robot is really cute, and having a voice conversation with it is quite interesting.'' (P15)}

\textit{``I need to be careful about what I say because I drive for DiDi, and there is voice monitoring in the car. If there is a robot like this, I would be more concerned about surveillance and privacy issues.'' (P24)}

It can be seen that with the progress and popularization of new technologies, people become aware of the risks and hazards behind them yet still lack a detailed and comprehensive understanding. Therefore, we conduct a second round of interview to prompt the participants to realize the existence of the risks. Before the follow-up interview, we first analyze what the participants said in the driving experiments to find potential risks expressed in their words. Hence, we mainly present these words to the participants and prompt them accordingly. An example is shown in Table~\ref{tab:prompt}.

\begin{table*}[ht]
  \caption{Prompt example of the follow-up interview with P23.}
  \label{tab:prompt}
  \begin{center}
  \begin{tabular}{l}
    \toprule
    \textit{\textbf{Answer from the initial interview}} \\
    \hdashline
    \textit{P23}: Concerns? What concerns? The danger of robots? I don't think it is as horrible as the Terminator haha. \\
    \midrule
    \textit{\textbf{Follow-up interview}} \\
    \hdashline
    \textit{Prompt}: When implementing the dialing task in the driving experiments, you said the name of ***. \\ \ \ \ \ \ \ \ \ \ \ \ \ \ \ Would you worry about the leakage of this information? \\
    \textit{P23}: \ \ \ \ \ \ Okay, it makes sense. Everything I said is basically at risk of being leaked. \\
    \textit{Prompt}: Exactly. Now, can you imagine other potential risks in the context of driving? \\
    \textit{P23}: \ \ \ \ \ \ My home address and the spots that I often visit can all be at risk. \\
    \textit{Prompt}: Perfect. How would these risks apply to other types of drivers, such as those of a different gender or age? \\ 
    \textit{P23}: \ \ \ \ \ \ If such information is leaked, then it is very dangerous for girls and the elderly who live and drive alone. \\
    \textit{Prompt}: Wonderful. Do the risks differ when communicating with a robot assistant? \\
    \textit{P23}: \ \ \ \ \ \ Let me think. (after a while) I feel that the cute appearance of this robot would stimulate my speaking \\ \ \ \ \ \ \ \ \ \ \ \ \ \ \ desire and let my guard down and unconsciously, I might reveal private details that I wasn't aware of. \\
    \bottomrule
  \end{tabular}
  \end{center}
\end{table*}

From the follow-up interview with P23, it is evident that, with our prompts, participants can transition from being completely unaware of potential risks to having a well-understood awareness. We applied similar prompts to other participants and obtained the following findings: \textbf{1)} Participants in the technology industry are more likely to be aware of privacy risks. \textbf{2)} Female participants are more conscious of the security risks associated with information leakage. \textbf{3)} Younger participants show greater sensitivity to the risks of surveillance. \textbf{4)} Interestingly, the appearance of the robot arouses the desire to speak, and its cute appearance can make people relax their guard and inadvertently disclose private information. Furthermore, one participant mentioned that VAs never seem to remind users to be cautious about privacy issues, and users easily perceive them as just cold machines, potentially ignoring any reminders. In contrast, with the appearance of a robot, such reminders might surprisingly be effective. Users may pay more attention to the warnings and alerts presented by the robot.

\section{Discussion}
\label{sec:discussion}
Based on all the above experimental findings, we discuss the advantages that a social robot like NOMI can bring to intelligent driving, as well as its shortcomings and the possibilities for expansion in its application areas.

\subsection{Voice Interaction}
From the user experiments, it is evident that the RA demonstrates satisfactory speech capabilities in voice interactions. Firstly, even without visual access to NOMI, users are generally content with the speech recognition and processing abilities. During driving, both the auditory workload and mental workload are relieved to some extent. However, some users reported that the RA's speech interaction abilities did not significantly differ from other VAs, especially the in-car ones. This is primarily due to two reasons. Firstly, at the speech level, the lack of variation in prosodic elements such as intonation and speaking rate, restricts the RA from conveying diverse emotions, attitudes, personalities, and pragmatic information \cite{ward2016interactional}. It fails to adapt to the varied requirements of different drivers based on gender, personality, and profession. These requirements are substantiated in linguistics, HCI, and psychology \cite{li2022robotic,edwards2019evaluations,nass2005wired,lee2000can,eyssel2012if}. Secondly, at the dialogue level, the VA lacks the initiative to initiate a conversation. In certain scenarios, users desire timely alerts similar to those provided by driving instructors, without running the risk of disrupting users who are already engaged in ongoing primary tasks \cite{meck2023may}. Additionally, the RA cannot retain dialogue memory or engage in multi-turn dialogue based on keywords from user speech, which has proven useful for increasing interactivity in various HCI scenarios \cite{misu2007speech,li2022alzheimer}. Finally, we believe there is a third reason: the RA's speech cannot adapt according to different scenarios. For instance, when executing user commands, the RA should respond quickly and accurately, yet in casual conversations, it could incorporate some human-like speech behaviors like fillers, backchannels, and self-corrections, which have proven to make dialogue systems more human-like \cite{li2019expressing,nakanishi2019generating,yamaguchi2016analysis,lala2017utterance}. We believe NOMI can alleviate some negative user experiences, but these limitations cannot be easily resolved by the robot NOMI.

\subsection{Multimodal Interaction}
From the comparative experiments, it is obvious that having NOMI significantly enhances the user experience both when stationary and while driving. Users highlighted three main benefits of the robot. Firstly, it encourages some users to engage in conversation, fostering a desire to interact, especially for long-term use. This indicates that multimodal cues, particularly visual signals, significantly enhance engagement. This aligns with previous findings in the field of human-robot interaction and digital content analysis \cite{inoue2021engagement,xia2021engaging}. Secondly, NOMI serves as a quiet, adorable pet for some users, effectively reducing the stress and workload of driving. Besides, it functions, to some extent, as a driving instructor or partner, providing reassurance in challenging driving scenarios, particularly for less experienced drivers. Thirdly, multiple users reflected that NOMI creates a feeling of companionship, which is considered its most significant value. In a word, the NOMI robot contributes a visual element that speech alone cannot provide. It features a circular display screen as a facial element, and offers a variety of expressions, creating diverse emotional displays to establish empathy.

\subsection{Emotional Interaction}
Given the user interest in NOMI offering emotional communication, we recognize it as an essential feature. Currently, the NOMI robot can convey several basic emotions through facial expressions, positively acknowledged by participants. We believe emotional interaction should progress based on precise user emotion recognition. In our prior work, we introduced an emotion engine that identified emotions from user speech \cite{li2021feeling}. However, this feature encountered challenges in broad implementation due to varied emotional expressions across demographics. For instance, female voices typically have higher frequencies than male voices. Determining emotions in such speech is complex without considering gender \cite{thellman2018he,manstead1992gender}.

To address this, we suggest integrating other modalities available in cars as supplements, as fusing multiple modalities has proven effective for emotion recognition \cite{wang2023cross}. For instance, the in-cabin cameras, often utilized to detect driver fatigue, can provide gender, age, and facial expression data, enhancing emotion recognition accuracy. Additionally, sensors for muscle signals on the steering wheel and for brainwave signals in the seat could further enhance emotion recognition through physiological indicators. Moreover, when initiating dialogues, the RA can start talking with emotion in specific scenarios to intrigue users' emotions to enhance emotional interaction \cite{li2023know,zepf2019towards}. This cannot only elevate human-car interaction but also alleviate driver fatigue and drowsiness. Last but not least, detecting emotion-related affective states would benefit the development of adaptive in-car systems based on different user profiles \cite{liao2022driver,wu2020driving}.

\subsection{Elderly Driving}
As NOMI can create a sense of `companionship' and `security,' and some participants felt it acted like an instructor, we believe the NOMI robot is suitable for elderly drivers. Firstly, previous research has explored the effectiveness of VAs for elderly driving. For instance, Alvarez \emph{et al.} \cite{alvarez2012voice} demonstrated that VAs can assist elderly drivers in looking for vehicle information with minimal impact on their driving performance, making them potentially safe applications as they do not significantly impair driving performance. Hong \cite{hong2015study} studied an adaptive assistant system for elderly driving and found that emotion adaptation is beneficial. Given the VA function and NOMI's facial emotion expression, it is expected that the RA can significantly improve the driving experience and safety for the elderly. Tanaka \emph{et al.} \cite{tanaka2020analysis} discovered that the safety of elderly driving can be increased with the assistance of a robot providing driving support and feedback.

Secondly, to enhance elderly drivers' skills and reduce accident rates, driving education systems are sometimes necessary. These systems are typically created using Virtual Reality (VR) technology in a highly immersive simulation scenario \cite{Luo2023vrcst, Seo2021vr}. While a wide variety of hazardous situations can be provided as training scenarios, the potential boredom of the training content may reduce the concentration of the elderly, and the expected training effect may not be achieved \cite{Luo2023vrcst}. As these systems do not support voice interaction or robot integration, we believe that an in-car assistant with a robot can facilitate more effective interaction, serving as a cutting-edge driving education platform. Moreover, by adapting the dialogue to the personality and emotions of elderly drivers, concentration can be sustained, thereby helping to improve the training effect.

\subsection{Autonomous Driving}
The findings we presented are all based on manual driving. However, in the context of autonomous driving, we believe the narrative is different. As drivers no longer have to focus much on driving tasks, in-car assistants should prioritize enhancing driving experiences by creating companionship and interaction. This is especially relevant in Level 5 autonomous driving, where drivers may be alone and bored in cars \cite{wang2021vehicle}. This represents a significant advantage that RAs can bring to the autonomous driving scenario.

More importantly, user trust takes precedence in autonomous driving compared to manual driving, as vehicles themselves assume more or even full control of the driving process, impacting road safety. Premstaller \emph{et al.} \cite{premstaller2023embodied} revealed that embodied conversational agents should support trust calibration by categorizing the in-car system into autonomous driving, infotainment (controlling music/media, calling contacts, climate control, etc.), and maintenance (system settings, service interval and gas level reminders, etc.). These different aspects require different agents with varying trust levels. Additionally, more voice commands that support diverse scenarios should be considered \cite{deruyttere2019talk2car}. 

\subsection{Phone-Based Versus Car-Based Assistants}
Due to some participants mentioned that the in-car features they commonly use can be achieved with a mobile phone, and noted that map updates on the phone are more timely, these are issues that in-car assistants need to address. Through experiments, we discovered that, in the presence of NOMI, users are more willing to interact with the RA. Since a physical robot is something a mobile phone cannot provide, we believe that one of the key focuses for the future development of in-car assistants is to emphasize the advantages of the robot and develop interactive features not available on mobile phones.

Regarding the issue of map updates, current in-car assistants can already address this through Over-The-Air (OTA) upgrades and built-in software updates, as well as vehicle connectivity features capable of sharing real-time traffic conditions \cite{shen2020cooperative}. However, hardware updates for in-car assistants are more challenging than for mobile phones. The replacement frequency of mobile phones is much higher than that of cars. We suggest that, in the future, all in-car VA functionalities could be integrated into the robot, rather than solely assigning robot the role of handling social interactions. Thus, directly replacing the robot might become a viable option in the event of an outdated chip. Furthermore, in the presence of the robot, participants exhibit higher tolerance for erroneous speech recognition results and high-latency feedback, which is not achievable with a mobile phone.

\subsection{Ethical, Legal, Privacy, and Safety Risks}

As we noted and discussed in Section~\ref{sec:concerns}, cameras and microphones may record activities and sensitive information of both drivers and passengers, potentially compromising personal privacy. Failure to secure this data adequately could lead to personal information leaks, privacy breaches, and misuse of data, resulting in tracking individual behavior, selling data to advertisers, or using it for other commercial purposes.

Additionally, biases may arise from algorithms employed in speech and language models, leading to discrimination against certain groups. Emerging research on responsible large and foundational speech and language models can be leveraged to mitigate such biases. For instance, the application of adversarial training in speech processing can safeguard gender information while preserving emotion information \cite{gao2023two}.

To mitigate these potential risks, it is essential to implement appropriate technological and regulatory measures to ensure data security, respect privacy, and minimize the risk of misuse. Additionally, clear regulations, legal and ethical guidelines should be established to encourage the industry to adopt responsible data processing and privacy protection measures.

\subsection{Limitations of Our Study}
This study has several limitations, and we would like to bring them to the forefront to emphasize their significance for future research. \textbf{1)} The findings in semi-supervised interviews, derived from a limited sample size, might not be easily generalized to the broader population of users. Notably, the risks associated with an in-car social robot may vary across countries due to inherent differences. As only Chinese participants participated in the non-driving and on-road experiments, we need a future work to include participants from other countries. \textbf{2)} Although we attempted to recruit participants from diverse ages, professions, places of residence, and balanced participants concerning gender, we did not consider their cultural backgrounds and driving ages which may influence participants’ views on the technologies involved. \textbf{3)} Due to the COVID-19 pandemic during this study, we were unable to conduct in-person interviews. This limitation might have affected participant engagement, possibly resulting in less comprehensive information. \textbf{4)} Given the rapid advancements of in-car VAs, it is important to note that some of our findings may not fully align with the capabilities of the most current systems. The NIO car's assistant, which we utilized for this study, represents its second version. The newest version has offered expanded functions, additional facial expressions for NOMI, and improved response times.

\section{Recommendations for the Community}
\label{sec:recommendations}

Designing in-car robot assistants requires a careful balance of functionality, distraction management, interaction dynamics, and regulation. We look into the following aspects and provide respective recommendations for the community.

\textbf{Aspect 1. Size and Mobility}

When considering in-car social robot design, the size is crucial for seamless integration with the limited vehicle space. A recommended approach is a compact and unobtrusive design that allows easy stowing or retraction when not in use. Unlike daily-use social robots, the focus should be on a practical form, minimizing physical presence to avoid obstructing the driver's view. Opting for limited and controlled movements, like subtle tilting or rotating, is preferable to prevent unnecessary troubles during driving.

\textbf{Aspect 2. Distraction Mitigation}

There has been research on voice-only assistants capable of reacting to signals from the environment to interrupt when the driver needs to be fully attentive to the driving task and subsequently resume voice delivery \cite{kennington2014better}. However, it is a challenging task in the context of robot assistants. To minimize potential distractions to drivers, we believe the design should prioritize hands-free and voice-controlled interactions. The robot's communication and movement need to be contextually aware, emphasizing providing information without diverting the driver's attention during critical driving moments. Incorporating informative yet unintrusive voice prompts or signals contributes to an overall safer driving experience.

\textbf{Aspect 3. Human-Robot Interaction Abilities}

The success of an in-car social robot relies on effective human-robot interaction. Natural language processing and understanding are crucial for facilitating seamless voice interactions. Implementing a user-friendly interface, possibly through a touchscreen or voice-activated commands, enhances accessibility and usability. Adaptive and context-aware responses contribute to an efficient and user-centric interaction model, aligning with the primary function of a voice assistant in the car.

\textbf{Aspect 4. Personalization, Customization, and Adaptation}

Recognizing the diverse preferences and needs of drivers, providing options for personalization and customization in the robot assistant's behavior and appearance can enhance the user experience. Tailoring the interactions to individual preferences contributes to a more personalized and engaging in-car assistant. Besides, considering the dynamic nature of driving conditions, the robot assistant should be adaptable to various situations. This includes adjusting its communication style (e.g., personality) based on traffic conditions, weather, and the driver's states, ensuring optimal interaction under different circumstances.

\textbf{Aspect 5. User Education and Onboarding}

As the robot assistant gains more trust, educating users about the capabilities and limitations of the robot assistant, as well as the vehicle systems, along with providing effective onboarding processes, can enhance user acceptance and satisfaction. This approach also helps reduce safety risks during subsequent driving. Clear communication on how to use and interact with the assistant promotes a positive user experience.

\textbf{Aspect 6. In-Depth Collaboration with Speech and Language Communities}

As foundation models are rapidly replacing traditional speech and language models, we believe they can be a game-changer in the automobile industry as well. The key lies in the smooth adaptation of foundation models in in-car scenarios, achieved through in-depth collaboration with the speech and language communities. Both sides need to understand each other well to build models, especially tailored for in-car scenarios.

\textbf{Aspect 7. Seeking Consultancy from Legal Experts}

The regulation of AI has intensified worldwide. As in-car RAs encompass many aspects of AI, including voice, language, vision, and more, such an integrated AI platform requires more regulation than single-function AI products. We believe it is necessary to seek consultancy from legal experts before designing the specifications of in-car RAs to address potential issues that certain functions might incur after being released onto the market.

\section{Conclusions}
\label{sec:conclusion}
In this comprehensive study, we delved into the effectiveness of an in-car assistant integrated with a social robot within real driving scenarios. The user study shed light on diverse user perspectives, particularly concerning in-car voice assistants. The results from both driving and non-driving experiments unveiled compelling insights: participants exhibited a greater willingness to engage in conversation and expressed heightened satisfaction when interacting with a voice assistant featuring a social robot. Moreover, they reported a sense of companionship and reassurance, particularly in challenging traffic conditions. The personality assessment indicated that the presence of a social robot contributed to the expression of salient personality traits. The inclusion of an open-ended question revealed concerns and risks related to ethical, legal, privacy, and safety issues associated with in-car robot assistants. 

Drawing from these results and findings, our study concludes with a set of recommendations. These suggestions are tailored for HCI researchers and automotive UX/UI designers, aiming to enhance engagement and overall driving experiences in the design of in-car assistants. Our goal is to drive initiatives that reinforce the standardization of data protection, privacy measures, and safety protocols. This two-fold strategy not only aims to enhance user experience but also addresses crucial concerns, creating a safer and more user-friendly environment for in-car assistant technologies. Furthermore, we aspire that the identified risks serve as a call to action, drawing the attention of relevant stakeholders.

\begin{acks}
Yuanchao would like to thank \textbf{Javier Hernandez} (Microsoft Research) for inspiration and discussions on the ethical issue study, and \textbf{Laura Weidinger} (Google DeepMind) for guidance and discussions on the experimental design of the interview study in Section~\ref{sec:concerns}. He appreciates the knowledge and experience gained while working at Honda Innovation Lab.
\end{acks}

\bibliographystyle{ACM-Reference-Format}
\bibliography{sample-base}


\begin{thebibliography}{88}


\ifx \showCODEN    \undefined \def \showCODEN     #1{\unskip}     \fi
\ifx \showDOI      \undefined \def \showDOI       #1{#1}\fi
\ifx \showISBNx    \undefined \def \showISBNx     #1{\unskip}     \fi
\ifx \showISBNxiii \undefined \def \showISBNxiii  #1{\unskip}     \fi
\ifx \showISSN     \undefined \def \showISSN      #1{\unskip}     \fi
\ifx \showLCCN     \undefined \def \showLCCN      #1{\unskip}     \fi
\ifx \shownote     \undefined \def \shownote      #1{#1}          \fi
\ifx \showarticletitle \undefined \def \showarticletitle #1{#1}   \fi
\ifx \showURL      \undefined \def \showURL       {\relax}        \fi
\providecommand\bibfield[2]{#2}
\providecommand\bibinfo[2]{#2}
\providecommand\natexlab[1]{#1}
\providecommand\showeprint[2][]{arXiv:#2}

\bibitem[\protect\citeauthoryear{Alvarez, L{\'o}pez-de Ipi{\~n}a, and
  Gilbert}{Alvarez et~al\mbox{.}}{2012}]%
        {alvarez2012voice}
\bibfield{author}{\bibinfo{person}{Ignacio Alvarez},
  \bibinfo{person}{Miren~Karmele L{\'o}pez-de Ipi{\~n}a}, {and}
  \bibinfo{person}{Juan~E Gilbert}.} \bibinfo{year}{2012}\natexlab{}.
\newblock \showarticletitle{The voice user help, a smart vehicle assistant for
  the elderly}. In \bibinfo{booktitle}{\emph{International Conference on
  Ubiquitous Computing and Ambient Intelligence}}. Springer,
  \bibinfo{pages}{314--321}.
\newblock


\bibitem[\protect\citeauthoryear{Amman, Huber, Charette, Richardson, and
  Wheeler}{Amman et~al\mbox{.}}{2017}]%
        {amman2017impact}
\bibfield{author}{\bibinfo{person}{Scott Amman}, \bibinfo{person}{John Huber},
  \bibinfo{person}{Francois Charette}, \bibinfo{person}{Brigitte Richardson},
  {and} \bibinfo{person}{Joshua Wheeler}.} \bibinfo{year}{2017}\natexlab{}.
\newblock \showarticletitle{The impact of microphone location and beamforming
  on in-vehicle speech recognition}.
\newblock \bibinfo{journal}{\emph{SAE International Journal of Passenger
  Cars-Electronic and Electrical Systems}} \bibinfo{volume}{10},
  \bibinfo{number}{2} (\bibinfo{year}{2017}).
\newblock


\bibitem[\protect\citeauthoryear{Braun, Mainz, Chadowitz, Pfleging, and
  Alt}{Braun et~al\mbox{.}}{2019}]%
        {braun2019your}
\bibfield{author}{\bibinfo{person}{Michael Braun}, \bibinfo{person}{Anja
  Mainz}, \bibinfo{person}{Ronee Chadowitz}, \bibinfo{person}{Bastian
  Pfleging}, {and} \bibinfo{person}{Florian Alt}.}
  \bibinfo{year}{2019}\natexlab{}.
\newblock \showarticletitle{At your service: Designing voice assistant
  personalities to improve automotive user interfaces}. In
  \bibinfo{booktitle}{\emph{Proceedings of the 2019 CHI Conference on Human
  Factors in Computing Systems}}. \bibinfo{pages}{1--11}.
\newblock


\bibitem[\protect\citeauthoryear{Cheng, Meng, Yao, and Wang}{Cheng
  et~al\mbox{.}}{2022}]%
        {cheng2022driving}
\bibfield{author}{\bibinfo{person}{Peiyao Cheng}, \bibinfo{person}{Fangang
  Meng}, \bibinfo{person}{Jie Yao}, {and} \bibinfo{person}{Yiran Wang}.}
  \bibinfo{year}{2022}\natexlab{}.
\newblock \showarticletitle{Driving With Agents: Investigating the Influences
  of Anthropomorphism Level and Physicality of Agents on Drivers' Perceived
  Control, Trust, and Driving Performance}.
\newblock \bibinfo{journal}{\emph{Frontiers in Psychology}}
  \bibinfo{volume}{13} (\bibinfo{year}{2022}).
\newblock


\bibitem[\protect\citeauthoryear{Cheng and Roedig}{Cheng and Roedig}{2022}]%
        {cheng2022personal}
\bibfield{author}{\bibinfo{person}{Peng Cheng} {and} \bibinfo{person}{Utz
  Roedig}.} \bibinfo{year}{2022}\natexlab{}.
\newblock \showarticletitle{Personal voice assistant security and privacy—a
  survey}.
\newblock \bibinfo{journal}{\emph{Proc. IEEE}} \bibinfo{volume}{110},
  \bibinfo{number}{4} (\bibinfo{year}{2022}), \bibinfo{pages}{476--507}.
\newblock


\bibitem[\protect\citeauthoryear{Christensen and Eriksen}{Christensen and
  Eriksen}{2021}]%
        {christensen2021your}
\bibfield{author}{\bibinfo{person}{Henrik Christensen} {and}
  \bibinfo{person}{Stian Eriksen}.} \bibinfo{year}{2021}\natexlab{}.
\newblock \emph{\bibinfo{title}{Is Your Voice Enough, Alexa? Assessing the Role
  of Digital Assistant Personality, Modality, and Product Involvement on
  Consumer Evaluations.}}
\newblock \bibinfo{thesistype}{Master's\ thesis}.
  \bibinfo{school}{Handelsh{\o}yskolen BI}.
\newblock


\bibitem[\protect\citeauthoryear{Cuadra, Li, Lee, Cho, and Ju}{Cuadra
  et~al\mbox{.}}{2021}]%
        {cuadra2021my}
\bibfield{author}{\bibinfo{person}{Andrea Cuadra}, \bibinfo{person}{Shuran Li},
  \bibinfo{person}{Hansol Lee}, \bibinfo{person}{Jason Cho}, {and}
  \bibinfo{person}{Wendy Ju}.} \bibinfo{year}{2021}\natexlab{}.
\newblock \showarticletitle{My bad! repairing intelligent voice assistant
  errors improves interaction}.
\newblock \bibinfo{journal}{\emph{Proceedings of the ACM on Human-Computer
  Interaction}} \bibinfo{volume}{5}, \bibinfo{number}{CSCW1}
  (\bibinfo{year}{2021}), \bibinfo{pages}{1--24}.
\newblock


\bibitem[\protect\citeauthoryear{Dabrowski and Munson}{Dabrowski and
  Munson}{2011}]%
        {dabrowski201140}
\bibfield{author}{\bibinfo{person}{Jim Dabrowski} {and}
  \bibinfo{person}{Ethan~V Munson}.} \bibinfo{year}{2011}\natexlab{}.
\newblock \showarticletitle{40 years of searching for the best computer system
  response time}.
\newblock \bibinfo{journal}{\emph{Interacting with Computers}}
  \bibinfo{volume}{23}, \bibinfo{number}{5} (\bibinfo{year}{2011}),
  \bibinfo{pages}{555--564}.
\newblock


\bibitem[\protect\citeauthoryear{De~Waard and Brookhuis}{De~Waard and
  Brookhuis}{1996}]%
        {de1996measurement}
\bibfield{author}{\bibinfo{person}{Dick De~Waard} {and} \bibinfo{person}{KA
  Brookhuis}.} \bibinfo{year}{1996}\natexlab{}.
\newblock \showarticletitle{The measurement of drivers' mental workload}.
\newblock  (\bibinfo{year}{1996}).
\newblock


\bibitem[\protect\citeauthoryear{Deruyttere, Vandenhende, Grujicic, Van~Gool,
  and Moens}{Deruyttere et~al\mbox{.}}{2019}]%
        {deruyttere2019talk2car}
\bibfield{author}{\bibinfo{person}{Thierry Deruyttere}, \bibinfo{person}{Simon
  Vandenhende}, \bibinfo{person}{Dusan Grujicic}, \bibinfo{person}{Luc
  Van~Gool}, {and} \bibinfo{person}{Marie-Francine Moens}.}
  \bibinfo{year}{2019}\natexlab{}.
\newblock \showarticletitle{Talk2car: Taking control of your self-driving car}.
\newblock \bibinfo{journal}{\emph{arXiv preprint arXiv:1909.10838}}
  (\bibinfo{year}{2019}).
\newblock


\bibitem[\protect\citeauthoryear{Edwards, Edwards, Stoll, Lin, and
  Massey}{Edwards et~al\mbox{.}}{2019}]%
        {edwards2019evaluations}
\bibfield{author}{\bibinfo{person}{Chad Edwards}, \bibinfo{person}{Autumn
  Edwards}, \bibinfo{person}{Brett Stoll}, \bibinfo{person}{Xialing Lin}, {and}
  \bibinfo{person}{Noelle Massey}.} \bibinfo{year}{2019}\natexlab{}.
\newblock \showarticletitle{Evaluations of an artificial intelligence
  instructor's voice: Social Identity Theory in human-robot interactions}.
\newblock \bibinfo{journal}{\emph{Computers in Human Behavior}}
  \bibinfo{volume}{90} (\bibinfo{year}{2019}), \bibinfo{pages}{357--362}.
\newblock


\bibitem[\protect\citeauthoryear{Eyssel, Kuchenbrandt, Bobinger, De~Ruiter, and
  Hegel}{Eyssel et~al\mbox{.}}{2012}]%
        {eyssel2012if}
\bibfield{author}{\bibinfo{person}{Friederike Eyssel}, \bibinfo{person}{Dieta
  Kuchenbrandt}, \bibinfo{person}{Simon Bobinger}, \bibinfo{person}{Laura
  De~Ruiter}, {and} \bibinfo{person}{Frank Hegel}.}
  \bibinfo{year}{2012}\natexlab{}.
\newblock \showarticletitle{'If you sound like me, you must be more human' on
  the interplay of robot and user features on human-robot acceptance and
  anthropomorphism}. In \bibinfo{booktitle}{\emph{Proceedings of the seventh
  annual ACM/IEEE international conference on Human-Robot Interaction}}.
  \bibinfo{pages}{125--126}.
\newblock


\bibitem[\protect\citeauthoryear{Fiske, Cuddy, Glick, and Xu}{Fiske
  et~al\mbox{.}}{2002}]%
        {fiske2002model}
\bibfield{author}{\bibinfo{person}{Susan~T Fiske}, \bibinfo{person}{Amy~JC
  Cuddy}, \bibinfo{person}{Peter Glick}, {and} \bibinfo{person}{Jun Xu}.}
  \bibinfo{year}{2002}\natexlab{}.
\newblock \showarticletitle{A model of (often mixed) stereotype content:
  competence and warmth respectively follow from perceived status and
  competition.}
\newblock \bibinfo{journal}{\emph{Journal of personality and social
  psychology}} \bibinfo{volume}{82}, \bibinfo{number}{6}
  (\bibinfo{year}{2002}), \bibinfo{pages}{878}.
\newblock


\bibitem[\protect\citeauthoryear{Fukui, Watanabe, and Kanazawa}{Fukui
  et~al\mbox{.}}{2018}]%
        {fukui2018sound}
\bibfield{author}{\bibinfo{person}{Masahiro Fukui}, \bibinfo{person}{Toshihiko
  Watanabe}, {and} \bibinfo{person}{Minato Kanazawa}.}
  \bibinfo{year}{2018}\natexlab{}.
\newblock \showarticletitle{Sound source separation for plural passenger speech
  recognition in smart mobility system}.
\newblock \bibinfo{journal}{\emph{IEEE Transactions on Consumer Electronics}}
  \bibinfo{volume}{64}, \bibinfo{number}{3} (\bibinfo{year}{2018}),
  \bibinfo{pages}{399--405}.
\newblock


\bibitem[\protect\citeauthoryear{Gao, Chu, and Kawahara}{Gao
  et~al\mbox{.}}{2023}]%
        {gao2023two}
\bibfield{author}{\bibinfo{person}{Yuan Gao}, \bibinfo{person}{Chenhui Chu},
  {and} \bibinfo{person}{Tatsuya Kawahara}.} \bibinfo{year}{2023}\natexlab{}.
\newblock \showarticletitle{Two-stage finetuning of wav2vec 2.0 for speech
  emotion recognition with ASR and gender pretraining}. In
  \bibinfo{booktitle}{\emph{Proc. Interspeech}}.
\newblock


\bibitem[\protect\citeauthoryear{Hergeth, Lorenz, Vilimek, and Krems}{Hergeth
  et~al\mbox{.}}{2016}]%
        {hergeth2016keep}
\bibfield{author}{\bibinfo{person}{Sebastian Hergeth}, \bibinfo{person}{Lutz
  Lorenz}, \bibinfo{person}{Roman Vilimek}, {and} \bibinfo{person}{Josef~F
  Krems}.} \bibinfo{year}{2016}\natexlab{}.
\newblock \showarticletitle{Keep your scanners peeled: Gaze behavior as a
  measure of automation trust during highly automated driving}.
\newblock \bibinfo{journal}{\emph{Human factors}} \bibinfo{volume}{58},
  \bibinfo{number}{3} (\bibinfo{year}{2016}), \bibinfo{pages}{509--519}.
\newblock


\bibitem[\protect\citeauthoryear{Hernandez, McDuff, Benavides, Amores, Maes,
  and Picard}{Hernandez et~al\mbox{.}}{2014}]%
        {hernandez2014autoemotive}
\bibfield{author}{\bibinfo{person}{Javier Hernandez}, \bibinfo{person}{Daniel
  McDuff}, \bibinfo{person}{Xavier Benavides}, \bibinfo{person}{Judith Amores},
  \bibinfo{person}{Pattie Maes}, {and} \bibinfo{person}{Rosalind Picard}.}
  \bibinfo{year}{2014}\natexlab{}.
\newblock \showarticletitle{AutoEmotive: bringing empathy to the driving
  experience to manage stress}.
\newblock In \bibinfo{booktitle}{\emph{Proceedings of the 2014 companion
  publication on Designing interactive systems}}. \bibinfo{pages}{53--56}.
\newblock


\bibitem[\protect\citeauthoryear{Hirsch and Pearce}{Hirsch and Pearce}{2000}]%
        {hirsch2000aurora}
\bibfield{author}{\bibinfo{person}{Hans-G{\"u}nter Hirsch} {and}
  \bibinfo{person}{David Pearce}.} \bibinfo{year}{2000}\natexlab{}.
\newblock \showarticletitle{The Aurora experimental framework for the
  performance evaluation of speech recognition systems under noisy conditions}.
  In \bibinfo{booktitle}{\emph{ASR2000-Automatic speech recognition: challenges
  for the new Millenium ISCA tutorial and research workshop (ITRW)}}.
\newblock


\bibitem[\protect\citeauthoryear{Hong}{Hong}{2015}]%
        {hong2015study}
\bibfield{author}{\bibinfo{person}{Seunghee Hong}.}
  \bibinfo{year}{2015}\natexlab{}.
\newblock \showarticletitle{Study on Adaptive Driving Assistant System for
  Elderly Drivers}.
\newblock  (\bibinfo{year}{2015}).
\newblock


\bibitem[\protect\citeauthoryear{Horswill and Plooy}{Horswill and
  Plooy}{2008}]%
        {horswill2008auditory}
\bibfield{author}{\bibinfo{person}{Mark~S Horswill} {and}
  \bibinfo{person}{Annaliese~M Plooy}.} \bibinfo{year}{2008}\natexlab{}.
\newblock \showarticletitle{Auditory feedback influences perceived driving
  speeds}.
\newblock \bibinfo{journal}{\emph{Perception}} \bibinfo{volume}{37},
  \bibinfo{number}{7} (\bibinfo{year}{2008}), \bibinfo{pages}{1037--1043}.
\newblock


\bibitem[\protect\citeauthoryear{Inoue, Lala, and Kawahara}{Inoue
  et~al\mbox{.}}{2022}]%
        {inoue2022can}
\bibfield{author}{\bibinfo{person}{Koji Inoue}, \bibinfo{person}{Divesh Lala},
  {and} \bibinfo{person}{Tatsuya Kawahara}.} \bibinfo{year}{2022}\natexlab{}.
\newblock \showarticletitle{Can a robot laugh with you?: Shared laughter
  generation for empathetic spoken dialogue}.
\newblock \bibinfo{journal}{\emph{Frontiers in Robotics and AI}}
  \bibinfo{volume}{9} (\bibinfo{year}{2022}), \bibinfo{pages}{234}.
\newblock


\bibitem[\protect\citeauthoryear{Inoue, Lala, Yamamoto, Takanashi, and
  Kawahara}{Inoue et~al\mbox{.}}{2021}]%
        {inoue2021engagement}
\bibfield{author}{\bibinfo{person}{Koji Inoue}, \bibinfo{person}{Divesh Lala},
  \bibinfo{person}{Kenta Yamamoto}, \bibinfo{person}{Katsuya Takanashi}, {and}
  \bibinfo{person}{Tatsuya Kawahara}.} \bibinfo{year}{2021}\natexlab{}.
\newblock \showarticletitle{Engagement-based adaptive behaviors for laboratory
  guide in human-robot dialogue}. In \bibinfo{booktitle}{\emph{Increasing
  Naturalness and Flexibility in Spoken Dialogue Interaction: 10th
  International Workshop on Spoken Dialogue Systems}}. Springer,
  \bibinfo{pages}{129--139}.
\newblock


\bibitem[\protect\citeauthoryear{John, Srivastava, et~al\mbox{.}}{John
  et~al\mbox{.}}{1999}]%
        {john1999big}
\bibfield{author}{\bibinfo{person}{Oliver~P John}, \bibinfo{person}{Sanjay
  Srivastava}, {et~al\mbox{.}}} \bibinfo{year}{1999}\natexlab{}.
\newblock \showarticletitle{The Big-Five trait taxonomy: History, measurement,
  and theoretical perspectives}.
\newblock  (\bibinfo{year}{1999}).
\newblock


\bibitem[\protect\citeauthoryear{Karatas, Tanaka, Yoshihara, Tanabe, Kojima,
  Endo, and Manabe}{Karatas et~al\mbox{.}}{2023}]%
        {karatas2023robotic}
\bibfield{author}{\bibinfo{person}{Nihan Karatas}, \bibinfo{person}{Takahiro
  Tanaka}, \bibinfo{person}{Yuki Yoshihara}, \bibinfo{person}{Hiroko Tanabe},
  \bibinfo{person}{Motoshi Kojima}, \bibinfo{person}{Masato Endo}, {and}
  \bibinfo{person}{Shuhei Manabe}.} \bibinfo{year}{2023}\natexlab{}.
\newblock \showarticletitle{Robotic-Human-Machine-Interface for Elderly
  Driving: Balancing Embodiment and Anthropomorphism for Improved Acceptance}.
  In \bibinfo{booktitle}{\emph{International Conference on Social Robotics}}.
  Springer, \bibinfo{pages}{240--253}.
\newblock


\bibitem[\protect\citeauthoryear{Karatas, Yoshikawa, and Okada}{Karatas
  et~al\mbox{.}}{2016}]%
        {karatas2016namida}
\bibfield{author}{\bibinfo{person}{Nihan Karatas}, \bibinfo{person}{Soshi
  Yoshikawa}, {and} \bibinfo{person}{Michio Okada}.}
  \bibinfo{year}{2016}\natexlab{}.
\newblock \showarticletitle{Namida: Sociable driving agents with multiparty
  conversation}. In \bibinfo{booktitle}{\emph{Proceedings of the Fourth
  International Conference on Human Agent Interaction}}.
  \bibinfo{pages}{35--42}.
\newblock


\bibitem[\protect\citeauthoryear{Kennington, Kousidis, Baumann, Buschmeier,
  Kopp, and Schlangen}{Kennington et~al\mbox{.}}{2014}]%
        {kennington2014better}
\bibfield{author}{\bibinfo{person}{Casey Kennington}, \bibinfo{person}{Spyros
  Kousidis}, \bibinfo{person}{Timo Baumann}, \bibinfo{person}{Hendrik
  Buschmeier}, \bibinfo{person}{Stefan Kopp}, {and} \bibinfo{person}{David
  Schlangen}.} \bibinfo{year}{2014}\natexlab{}.
\newblock \showarticletitle{Better driving and recall when in-car information
  presentation uses situationally-aware incremental speech output generation}.
  In \bibinfo{booktitle}{\emph{Proceedings of the 6th international conference
  on automotive user interfaces and interactive vehicular applications}}.
  \bibinfo{pages}{1--7}.
\newblock


\bibitem[\protect\citeauthoryear{Kim and Heo}{Kim and Heo}{2021}]%
        {kim2021please}
\bibfield{author}{\bibinfo{person}{Jongkeon Kim} {and}
  \bibinfo{person}{Jeongyun Heo}.} \bibinfo{year}{2021}\natexlab{}.
\newblock \showarticletitle{Please stop listening while i make a private call:
  Context-aware in-vehicle mode of a voice-controlled intelligent personal
  assistant with a privacy consideration}. In
  \bibinfo{booktitle}{\emph{International Conference on Human-Computer
  Interaction}}. Springer, \bibinfo{pages}{177--193}.
\newblock


\bibitem[\protect\citeauthoryear{Kim, Park, and Park}{Kim
  et~al\mbox{.}}{2000}]%
        {kim2000psychophysiological}
\bibfield{author}{\bibinfo{person}{Jung-Yong Kim}, \bibinfo{person}{Min-Yong
  Park}, {and} \bibinfo{person}{Chang-Soon Park}.}
  \bibinfo{year}{2000}\natexlab{}.
\newblock \showarticletitle{Psychophysiological Responses Reflecting Driver's
  Emotional Reaction to Interior Noise during Sumulated Driving}. In
  \bibinfo{booktitle}{\emph{Proceedings of the Human Factors and Ergonomics
  Society Annual Meeting}}, Vol.~\bibinfo{volume}{44}. SAGE Publications Sage
  CA: Los Angeles, CA, \bibinfo{pages}{3--196}.
\newblock


\bibitem[\protect\citeauthoryear{Komatsu, Kurosawa, and Yamada}{Komatsu
  et~al\mbox{.}}{2012}]%
        {komatsu2012does}
\bibfield{author}{\bibinfo{person}{Takanori Komatsu}, \bibinfo{person}{Rie
  Kurosawa}, {and} \bibinfo{person}{Seiji Yamada}.}
  \bibinfo{year}{2012}\natexlab{}.
\newblock \showarticletitle{How Does the Difference Between Users’
  Expectations and Perceptions About a Robotic Agent Affect Their Behavior? An
  Adaptation Gap Concept for Determining Whether Interactions Between Users and
  Agents Are Going Well or Not}.
\newblock \bibinfo{journal}{\emph{International Journal of Social Robotics}}
  \bibinfo{volume}{4} (\bibinfo{year}{2012}), \bibinfo{pages}{109--116}.
\newblock


\bibitem[\protect\citeauthoryear{Lala, Li, and Kawahara}{Lala
  et~al\mbox{.}}{2017}]%
        {lala2017utterance}
\bibfield{author}{\bibinfo{person}{Divesh Lala}, \bibinfo{person}{Yuanchao Li},
  {and} \bibinfo{person}{Tatsuya Kawahara}.} \bibinfo{year}{2017}\natexlab{}.
\newblock \showarticletitle{Utterance Behavior of Users While Playing
  Basketball with a Virtual Teammate.}. In \bibinfo{booktitle}{\emph{ICAART
  (1)}}. \bibinfo{pages}{28--38}.
\newblock


\bibitem[\protect\citeauthoryear{Laugwitz, Held, and Schrepp}{Laugwitz
  et~al\mbox{.}}{2008}]%
        {laugwitz2008construction}
\bibfield{author}{\bibinfo{person}{Bettina Laugwitz}, \bibinfo{person}{Theo
  Held}, {and} \bibinfo{person}{Martin Schrepp}.}
  \bibinfo{year}{2008}\natexlab{}.
\newblock \showarticletitle{Construction and evaluation of a user experience
  questionnaire}. In \bibinfo{booktitle}{\emph{HCI and Usability for Education
  and Work: 4th Symposium of the Workgroup Human-Computer Interaction and
  Usability Engineering of the Austrian Computer Society, USAB 2008, Graz,
  Austria, November 20-21, 2008. Proceedings 4}}. Springer,
  \bibinfo{pages}{63--76}.
\newblock


\bibitem[\protect\citeauthoryear{Lauridsen, Gimenez, Rodriguez, Sorensen, and
  Mogensen}{Lauridsen et~al\mbox{.}}{2017}]%
        {lauridsen2017lte}
\bibfield{author}{\bibinfo{person}{Mads Lauridsen},
  \bibinfo{person}{Lucas~Chavarria Gimenez}, \bibinfo{person}{Ignacio
  Rodriguez}, \bibinfo{person}{Troels~B Sorensen}, {and}
  \bibinfo{person}{Preben Mogensen}.} \bibinfo{year}{2017}\natexlab{}.
\newblock \showarticletitle{From LTE to 5G for connected mobility}.
\newblock \bibinfo{journal}{\emph{IEEE Communications Magazine}}
  \bibinfo{volume}{55}, \bibinfo{number}{3} (\bibinfo{year}{2017}),
  \bibinfo{pages}{156--162}.
\newblock


\bibitem[\protect\citeauthoryear{Lee, Nass, and Brave}{Lee
  et~al\mbox{.}}{2000}]%
        {lee2000can}
\bibfield{author}{\bibinfo{person}{Eun~Ju Lee}, \bibinfo{person}{Clifford
  Nass}, {and} \bibinfo{person}{Scott Brave}.} \bibinfo{year}{2000}\natexlab{}.
\newblock \showarticletitle{Can computer-generated speech have gender? An
  experimental test of gender stereotype}. In \bibinfo{booktitle}{\emph{CHI'00
  extended abstracts on Human factors in computing systems}}.
  \bibinfo{pages}{289--290}.
\newblock


\bibitem[\protect\citeauthoryear{Lee, Caven, Haake, and Brown}{Lee
  et~al\mbox{.}}{2001}]%
        {lee2001speech}
\bibfield{author}{\bibinfo{person}{John~D Lee}, \bibinfo{person}{Brent Caven},
  \bibinfo{person}{Steven Haake}, {and} \bibinfo{person}{Timothy~L Brown}.}
  \bibinfo{year}{2001}\natexlab{}.
\newblock \showarticletitle{Speech-based interaction with in-vehicle computers:
  The effect of speech-based e-mail on drivers' attention to the roadway}.
\newblock \bibinfo{journal}{\emph{Human factors}} \bibinfo{volume}{43},
  \bibinfo{number}{4} (\bibinfo{year}{2001}), \bibinfo{pages}{631--640}.
\newblock


\bibitem[\protect\citeauthoryear{Li}{Li}{2018}]%
        {li2018towards}
\bibfield{author}{\bibinfo{person}{Yuanchao Li}.}
  \bibinfo{year}{2018}\natexlab{}.
\newblock \showarticletitle{Towards Improving Speech Emotion Recognition for
  In-Vehicle Agents: Preliminary Results of Incorporating Sentiment Analysis by
  Using Early and Late Fusion Methods}. In
  \bibinfo{booktitle}{\emph{Proceedings of the 6th International Conference on
  Human-Agent Interaction}}. \bibinfo{pages}{365--367}.
\newblock


\bibitem[\protect\citeauthoryear{Li}{Li}{2021}]%
        {li2021feeling}
\bibfield{author}{\bibinfo{person}{Yuanchao Li}.}
  \bibinfo{year}{2021}\natexlab{}.
\newblock \bibinfo{title}{Feeling estimation device, feeling estimation method,
  and storage medium}.
\newblock
\newblock
\newblock
\shownote{US Patent 11,107,464}.


\bibitem[\protect\citeauthoryear{Li, Inoue, Tian, Fu, Ishi, Ishiguro, Kawahara,
  and Lai}{Li et~al\mbox{.}}{2023a}]%
        {li2023know}
\bibfield{author}{\bibinfo{person}{Yuanchao Li}, \bibinfo{person}{Koji Inoue},
  \bibinfo{person}{Leimin Tian}, \bibinfo{person}{Changzeng Fu},
  \bibinfo{person}{Carlos~Toshinori Ishi}, \bibinfo{person}{Hiroshi Ishiguro},
  \bibinfo{person}{Tatsuya Kawahara}, {and} \bibinfo{person}{Catherine Lai}.}
  \bibinfo{year}{2023}\natexlab{a}.
\newblock \showarticletitle{I Know Your Feelings Before You Do: Predicting
  Future Affective Reactions in Human-Computer Dialogue}. In
  \bibinfo{booktitle}{\emph{Extended Abstracts of the 2023 CHI Conference on
  Human Factors in Computing Systems}}. \bibinfo{pages}{1--7}.
\newblock


\bibitem[\protect\citeauthoryear{Li, Ishi, Inoue, Nakamura, and Kawahara}{Li
  et~al\mbox{.}}{2019}]%
        {li2019expressing}
\bibfield{author}{\bibinfo{person}{Yuanchao Li},
  \bibinfo{person}{Carlos~Toshinori Ishi}, \bibinfo{person}{Koji Inoue},
  \bibinfo{person}{Shizuka Nakamura}, {and} \bibinfo{person}{Tatsuya
  Kawahara}.} \bibinfo{year}{2019}\natexlab{}.
\newblock \showarticletitle{Expressing reactive emotion based on multimodal
  emotion recognition for natural conversation in human--robot interaction}.
\newblock \bibinfo{journal}{\emph{Advanced Robotics}} \bibinfo{volume}{33},
  \bibinfo{number}{20} (\bibinfo{year}{2019}), \bibinfo{pages}{1030--1041}.
\newblock


\bibitem[\protect\citeauthoryear{Li and Lai}{Li and Lai}{2022}]%
        {li2022robotic}
\bibfield{author}{\bibinfo{person}{Yuanchao Li} {and}
  \bibinfo{person}{Catherine Lai}.} \bibinfo{year}{2022}\natexlab{}.
\newblock \showarticletitle{Robotic Speech Synthesis: Perspectives on
  Interactions, Scenarios, and Ethics}.
\newblock \bibinfo{journal}{\emph{2022 17th ACM/IEEE International Conference
  on Human-Robot Interaction (HRI)}} (\bibinfo{year}{2022}).
\newblock


\bibitem[\protect\citeauthoryear{Li, Lai, Lala, Inoue, and Kawahara}{Li
  et~al\mbox{.}}{2022}]%
        {li2022alzheimer}
\bibfield{author}{\bibinfo{person}{Yuanchao Li}, \bibinfo{person}{Catherine
  Lai}, \bibinfo{person}{Divesh Lala}, \bibinfo{person}{Koji Inoue}, {and}
  \bibinfo{person}{Tatsuya Kawahara}.} \bibinfo{year}{2022}\natexlab{}.
\newblock \showarticletitle{Alzheimer's Dementia Detection through Spontaneous
  Dialogue with Proactive Robotic Listeners}. In \bibinfo{booktitle}{\emph{2022
  17th ACM/IEEE International Conference on Human-Robot Interaction (HRI)}}.
  IEEE, \bibinfo{pages}{875--879}.
\newblock


\bibitem[\protect\citeauthoryear{Li, Zhao, Klejch, Bell, and Lai}{Li
  et~al\mbox{.}}{2023b}]%
        {li2023asr}
\bibfield{author}{\bibinfo{person}{Yuanchao Li}, \bibinfo{person}{Zeyu Zhao},
  \bibinfo{person}{Ondrej Klejch}, \bibinfo{person}{Peter Bell}, {and}
  \bibinfo{person}{Catherine Lai}.} \bibinfo{year}{2023}\natexlab{b}.
\newblock \showarticletitle{ASR and Emotional Speech: A Word-Level
  Investigation of the Mutual Impact of Speech and Emotion Recognition}.
\newblock \bibinfo{journal}{\emph{INTERSPEECH 2023}} (\bibinfo{year}{2023}).
\newblock


\bibitem[\protect\citeauthoryear{Liao, Mehrotra, Ho, Gorospe, Wu, and
  Mistu}{Liao et~al\mbox{.}}{2022}]%
        {liao2022driver}
\bibfield{author}{\bibinfo{person}{Xishun Liao}, \bibinfo{person}{Shashank
  Mehrotra}, \bibinfo{person}{Samson Ho}, \bibinfo{person}{Yuki Gorospe},
  \bibinfo{person}{Xingwei Wu}, {and} \bibinfo{person}{Teruhisa Mistu}.}
  \bibinfo{year}{2022}\natexlab{}.
\newblock \showarticletitle{Driver profile modeling based on driving style,
  personality traits, and mood states}. In \bibinfo{booktitle}{\emph{2022 IEEE
  25th international conference on intelligent transportation systems (ITSC)}}.
  IEEE, \bibinfo{pages}{709--716}.
\newblock


\bibitem[\protect\citeauthoryear{Liu, Wan, Zou, and Zhang}{Liu
  et~al\mbox{.}}{2023}]%
        {liu2023exploring}
\bibfield{author}{\bibinfo{person}{Jing Liu}, \bibinfo{person}{Fucheng Wan},
  \bibinfo{person}{Jinzhi Zou}, {and} \bibinfo{person}{Jiaqi Zhang}.}
  \bibinfo{year}{2023}\natexlab{}.
\newblock \showarticletitle{Exploring Factors Affecting People’s Willingness
  to Use a Voice-Based In-Car Assistant in Electric Cars: An Empirical Study}.
\newblock \bibinfo{journal}{\emph{World Electric Vehicle Journal}}
  \bibinfo{volume}{14}, \bibinfo{number}{3} (\bibinfo{year}{2023}),
  \bibinfo{pages}{73}.
\newblock


\bibitem[\protect\citeauthoryear{Lo and Green}{Lo and Green}{2013}]%
        {lo2013development}
\bibfield{author}{\bibinfo{person}{Victor Ei-Wen Lo} {and}
  \bibinfo{person}{Paul~A Green}.} \bibinfo{year}{2013}\natexlab{}.
\newblock \showarticletitle{Development and evaluation of automotive speech
  interfaces: useful information from the human factors and the related
  literature}.
\newblock \bibinfo{journal}{\emph{International Journal of Vehicular
  Technology}}  \bibinfo{volume}{2013} (\bibinfo{year}{2013}).
\newblock


\bibitem[\protect\citeauthoryear{Lopatovska}{Lopatovska}{2020}]%
        {lopatovska2020personality}
\bibfield{author}{\bibinfo{person}{Irene Lopatovska}.}
  \bibinfo{year}{2020}\natexlab{}.
\newblock \showarticletitle{Personality dimensions of intelligent personal
  assistants}. In \bibinfo{booktitle}{\emph{Proceedings of the 2020 conference
  on human information interaction and retrieval}}. \bibinfo{pages}{333--337}.
\newblock


\bibitem[\protect\citeauthoryear{Luo, Ahn, Abbas, Seo, Cha, and Kim}{Luo
  et~al\mbox{.}}{2023}]%
        {Luo2023vrcst}
\bibfield{author}{\bibinfo{person}{Yanfang Luo}, \bibinfo{person}{Seungahn
  Ahn}, \bibinfo{person}{Ali Abbas}, \bibinfo{person}{Joonoh Seo},
  \bibinfo{person}{Seunghyun Cha}, {and} \bibinfo{person}{Jungin Kim}.}
  \bibinfo{year}{2023}\natexlab{}.
\newblock \showarticletitle{Investigating the Impact of Scenario and
  Interaction Fidelity on Training Experience When Designing Immersive Virtual
  Reality-based Construction Safety Training.}
\newblock \bibinfo{journal}{\emph{Developments in the Built Environment}}
  \bibinfo{volume}{16} (\bibinfo{year}{2023}), \bibinfo{pages}{100223}.
\newblock


\bibitem[\protect\citeauthoryear{Manstead}{Manstead}{1992}]%
        {manstead1992gender}
\bibfield{author}{\bibinfo{person}{Antony~SR Manstead}.}
  \bibinfo{year}{1992}\natexlab{}.
\newblock \showarticletitle{Gender differences in emotion.}
\newblock  (\bibinfo{year}{1992}).
\newblock


\bibitem[\protect\citeauthoryear{Meck, Draxler, and Vogt}{Meck
  et~al\mbox{.}}{2023}]%
        {meck2023may}
\bibfield{author}{\bibinfo{person}{Anna-Maria Meck}, \bibinfo{person}{Christoph
  Draxler}, {and} \bibinfo{person}{Thurid Vogt}.}
  \bibinfo{year}{2023}\natexlab{}.
\newblock \showarticletitle{How May I Interrupt? Linguistic-Driven Design
  Guidelines for Proactive in-Car Voice Assistants}.
\newblock \bibinfo{journal}{\emph{International Journal of Human--Computer
  Interaction}} (\bibinfo{year}{2023}), \bibinfo{pages}{1--15}.
\newblock


\bibitem[\protect\citeauthoryear{Meyer, Shinar, Bitan, and Leiser}{Meyer
  et~al\mbox{.}}{1996}]%
        {meyer1996duration}
\bibfield{author}{\bibinfo{person}{Joachim Meyer}, \bibinfo{person}{David
  Shinar}, \bibinfo{person}{Yuval Bitan}, {and} \bibinfo{person}{David
  Leiser}.} \bibinfo{year}{1996}\natexlab{}.
\newblock \showarticletitle{Duration estimates and users' preferences in
  human-computer interaction}.
\newblock \bibinfo{journal}{\emph{Ergonomics}} \bibinfo{volume}{39},
  \bibinfo{number}{1} (\bibinfo{year}{1996}), \bibinfo{pages}{46--60}.
\newblock


\bibitem[\protect\citeauthoryear{Misu and Kawahara}{Misu and Kawahara}{2007}]%
        {misu2007speech}
\bibfield{author}{\bibinfo{person}{Teruhisa Misu} {and}
  \bibinfo{person}{Tatsuya Kawahara}.} \bibinfo{year}{2007}\natexlab{}.
\newblock \showarticletitle{Speech-based interactive information guidance
  system using question-answering technique}. In \bibinfo{booktitle}{\emph{2007
  IEEE International Conference on Acoustics, Speech and Signal
  Processing-ICASSP'07}}, Vol.~\bibinfo{volume}{4}. IEEE,
  \bibinfo{pages}{IV--145}.
\newblock


\bibitem[\protect\citeauthoryear{Mitra, Huang, Lea, Tooley, Wu, Botten,
  Palekar, Thelapurath, Georgiou, Kajarekar, et~al\mbox{.}}{Mitra
  et~al\mbox{.}}{2021}]%
        {mitra2021analysis}
\bibfield{author}{\bibinfo{person}{Vikramjit Mitra}, \bibinfo{person}{Zifang
  Huang}, \bibinfo{person}{Colin Lea}, \bibinfo{person}{Lauren Tooley},
  \bibinfo{person}{Sarah Wu}, \bibinfo{person}{Darren Botten},
  \bibinfo{person}{Ashwini Palekar}, \bibinfo{person}{Shrinath Thelapurath},
  \bibinfo{person}{Panayiotis Georgiou}, \bibinfo{person}{Sachin Kajarekar},
  {et~al\mbox{.}}} \bibinfo{year}{2021}\natexlab{}.
\newblock \showarticletitle{Analysis and tuning of a voice assistant system for
  dysfluent speech}.
\newblock \bibinfo{journal}{\emph{arXiv preprint arXiv:2106.11759}}
  (\bibinfo{year}{2021}).
\newblock


\bibitem[\protect\citeauthoryear{Nafari and Weaver}{Nafari and Weaver}{2013}]%
        {nafari2013augmenting}
\bibfield{author}{\bibinfo{person}{Maryam Nafari} {and} \bibinfo{person}{Chris
  Weaver}.} \bibinfo{year}{2013}\natexlab{}.
\newblock \showarticletitle{Augmenting visualization with natural language
  translation of interaction: A usability study}. In
  \bibinfo{booktitle}{\emph{Computer Graphics Forum}},
  Vol.~\bibinfo{volume}{32}. Wiley Online Library, \bibinfo{pages}{391--400}.
\newblock


\bibitem[\protect\citeauthoryear{Nakanishi, Inoue, Nakamura, Takanashi, and
  Kawahara}{Nakanishi et~al\mbox{.}}{2019}]%
        {nakanishi2019generating}
\bibfield{author}{\bibinfo{person}{Ryosuke Nakanishi}, \bibinfo{person}{Koji
  Inoue}, \bibinfo{person}{Shizuka Nakamura}, \bibinfo{person}{Katsuya
  Takanashi}, {and} \bibinfo{person}{Tatsuya Kawahara}.}
  \bibinfo{year}{2019}\natexlab{}.
\newblock \showarticletitle{Generating fillers based on dialog act pairs for
  smooth turn-taking by humanoid robot}. In \bibinfo{booktitle}{\emph{9th
  International Workshop on Spoken Dialogue System Technology}}. Springer,
  \bibinfo{pages}{91--101}.
\newblock


\bibitem[\protect\citeauthoryear{Nass and Brave}{Nass and Brave}{2005}]%
        {nass2005wired}
\bibfield{author}{\bibinfo{person}{Clifford~Ivar Nass} {and}
  \bibinfo{person}{Scott Brave}.} \bibinfo{year}{2005}\natexlab{}.
\newblock \bibinfo{booktitle}{\emph{Wired for speech: How voice activates and
  advances the human-computer relationship}}.
\newblock \bibinfo{publisher}{MIT press Cambridge}.
\newblock


\bibitem[\protect\citeauthoryear{Noll}{Noll}{1990}]%
        {noll1990problems}
\bibfield{author}{\bibinfo{person}{A Noll}.} \bibinfo{year}{1990}\natexlab{}.
\newblock \showarticletitle{Problems of speech recognition in mobile
  environments.}. In \bibinfo{booktitle}{\emph{ICSLP}}.
  \bibinfo{pages}{1133--1136}.
\newblock


\bibitem[\protect\citeauthoryear{O'Brien and DeLongis}{O'Brien and
  DeLongis}{1996}]%
        {o1996interactional}
\bibfield{author}{\bibinfo{person}{Tess~Byrd O'Brien} {and}
  \bibinfo{person}{Anita DeLongis}.} \bibinfo{year}{1996}\natexlab{}.
\newblock \showarticletitle{The interactional context of problem-, emotion-,
  and relationship-focused coping: the role of the big five personality
  factors}.
\newblock \bibinfo{journal}{\emph{Journal of personality}}
  \bibinfo{volume}{64}, \bibinfo{number}{4} (\bibinfo{year}{1996}),
  \bibinfo{pages}{775--813}.
\newblock


\bibitem[\protect\citeauthoryear{Pauzi{\'e}}{Pauzi{\'e}}{2008}]%
        {pauzie2008method}
\bibfield{author}{\bibinfo{person}{Annie Pauzi{\'e}}.}
  \bibinfo{year}{2008}\natexlab{}.
\newblock \showarticletitle{A method to assess the driver mental workload: The
  driving activity load index (DALI)}.
\newblock \bibinfo{journal}{\emph{IET Intelligent Transport Systems}}
  \bibinfo{volume}{2}, \bibinfo{number}{4} (\bibinfo{year}{2008}),
  \bibinfo{pages}{315--322}.
\newblock


\bibitem[\protect\citeauthoryear{Paxion, Galy, and Berthelon}{Paxion
  et~al\mbox{.}}{2014}]%
        {paxion2014mental}
\bibfield{author}{\bibinfo{person}{Julie Paxion}, \bibinfo{person}{Edith Galy},
  {and} \bibinfo{person}{Catherine Berthelon}.}
  \bibinfo{year}{2014}\natexlab{}.
\newblock \showarticletitle{Mental workload and driving}.
\newblock \bibinfo{journal}{\emph{Frontiers in psychology}}
  \bibinfo{volume}{5} (\bibinfo{year}{2014}), \bibinfo{pages}{1344}.
\newblock


\bibitem[\protect\citeauthoryear{Pollmann, Ruff, Vetter, and
  Zimmermann}{Pollmann et~al\mbox{.}}{2020}]%
        {pollmann2020robot}
\bibfield{author}{\bibinfo{person}{Kathrin Pollmann},
  \bibinfo{person}{Christopher Ruff}, \bibinfo{person}{Kevin Vetter}, {and}
  \bibinfo{person}{Gottfried Zimmermann}.} \bibinfo{year}{2020}\natexlab{}.
\newblock \showarticletitle{Robot vs. voice assistant: Is playing with pepper
  more fun than playing with alexa?}. In \bibinfo{booktitle}{\emph{Companion of
  the 2020 ACM/IEEE international conference on human-robot interaction}}.
  \bibinfo{pages}{395--397}.
\newblock


\bibitem[\protect\citeauthoryear{Portet, Vacher, Golanski, Roux, and
  Meillon}{Portet et~al\mbox{.}}{2013}]%
        {portet2013design}
\bibfield{author}{\bibinfo{person}{Fran{\c{c}}ois Portet},
  \bibinfo{person}{Michel Vacher}, \bibinfo{person}{Caroline Golanski},
  \bibinfo{person}{Camille Roux}, {and} \bibinfo{person}{Brigitte Meillon}.}
  \bibinfo{year}{2013}\natexlab{}.
\newblock \showarticletitle{Design and evaluation of a smart home voice
  interface for the elderly: acceptability and objection aspects}.
\newblock \bibinfo{journal}{\emph{Personal and Ubiquitous Computing}}
  \bibinfo{volume}{17}, \bibinfo{number}{1} (\bibinfo{year}{2013}),
  \bibinfo{pages}{127--144}.
\newblock


\bibitem[\protect\citeauthoryear{Poushneh}{Poushneh}{2021}]%
        {poushneh2021humanizing}
\bibfield{author}{\bibinfo{person}{Atieh Poushneh}.}
  \bibinfo{year}{2021}\natexlab{}.
\newblock \showarticletitle{Humanizing voice assistant: The impact of voice
  assistant personality on consumers’ attitudes and behaviors}.
\newblock \bibinfo{journal}{\emph{Journal of Retailing and Consumer Services}}
  \bibinfo{volume}{58} (\bibinfo{year}{2021}), \bibinfo{pages}{102283}.
\newblock


\bibitem[\protect\citeauthoryear{Premstaller, Kotsios, and
  Wintersberger}{Premstaller et~al\mbox{.}}{2023}]%
        {premstaller2023embodied}
\bibfield{author}{\bibinfo{person}{Marie Premstaller}, \bibinfo{person}{Heike
  Kotsios}, {and} \bibinfo{person}{Philipp Wintersberger}.}
  \bibinfo{year}{2023}\natexlab{}.
\newblock \showarticletitle{Embodied Conversational Agent Teams for Trust
  Calibration in Automated Vehicles}. In \bibinfo{booktitle}{\emph{Adjunct
  Proceedings of the 15th International Conference on Automotive User
  Interfaces and Interactive Vehicular Applications}}. \bibinfo{pages}{71--76}.
\newblock


\bibitem[\protect\citeauthoryear{Riek, Paul, and Robinson}{Riek
  et~al\mbox{.}}{2010}]%
        {riek2010my}
\bibfield{author}{\bibinfo{person}{Laurel~D Riek}, \bibinfo{person}{Philip~C
  Paul}, {and} \bibinfo{person}{Peter Robinson}.}
  \bibinfo{year}{2010}\natexlab{}.
\newblock \showarticletitle{When my robot smiles at me: Enabling human-robot
  rapport via real-time head gesture mimicry}.
\newblock \bibinfo{journal}{\emph{Journal on Multimodal User Interfaces}}
  \bibinfo{volume}{3} (\bibinfo{year}{2010}), \bibinfo{pages}{99--108}.
\newblock


\bibitem[\protect\citeauthoryear{Schmidt, Minker, and Werner}{Schmidt
  et~al\mbox{.}}{2020}]%
        {schmidt2020users}
\bibfield{author}{\bibinfo{person}{Maria Schmidt}, \bibinfo{person}{Wolfgang
  Minker}, {and} \bibinfo{person}{Steffen Werner}.}
  \bibinfo{year}{2020}\natexlab{}.
\newblock \showarticletitle{How users react to proactive voice assistant
  behavior while driving}. In \bibinfo{booktitle}{\emph{Proceedings of The 12th
  Language Resources and Evaluation Conference}}. \bibinfo{pages}{485--490}.
\newblock


\bibitem[\protect\citeauthoryear{Seo, Park, Son, and Hong}{Seo
  et~al\mbox{.}}{2021}]%
        {Seo2021vr}
\bibfield{author}{\bibinfo{person}{Hyunjeong Seo}, \bibinfo{person}{Gyumi
  Park}, \bibinfo{person}{Minjie Son}, {and} \bibinfo{person}{Ahjeong Hong}.}
  \bibinfo{year}{2021}\natexlab{}.
\newblock \showarticletitle{Establishment of Virtual-Reality-Based Safety
  Education and Training System for Safety Engagement.}
\newblock \bibinfo{journal}{\emph{Education Sciences}} \bibinfo{volume}{11},
  \bibinfo{number}{12} (\bibinfo{year}{2021}), \bibinfo{pages}{786}.
\newblock


\bibitem[\protect\citeauthoryear{Shen, Zhang, Ouyang, Li, and
  Raksincharoensak}{Shen et~al\mbox{.}}{2020}]%
        {shen2020cooperative}
\bibfield{author}{\bibinfo{person}{Xun Shen}, \bibinfo{person}{Xingguo Zhang},
  \bibinfo{person}{Tinghui Ouyang}, \bibinfo{person}{Yuanchao Li}, {and}
  \bibinfo{person}{Pongsathorn Raksincharoensak}.}
  \bibinfo{year}{2020}\natexlab{}.
\newblock \showarticletitle{Cooperative Comfortable-Driving at Signalized
  Intersections for Connected and Automated Vehicles}.
\newblock \bibinfo{journal}{\emph{IEEE Robotics and Automation Letters}}
  \bibinfo{volume}{5}, \bibinfo{number}{4} (\bibinfo{year}{2020}),
  \bibinfo{pages}{6247--6254}.
\newblock


\bibitem[\protect\citeauthoryear{Shneiderman}{Shneiderman}{1984}]%
        {shneiderman1984response}
\bibfield{author}{\bibinfo{person}{Ben Shneiderman}.}
  \bibinfo{year}{1984}\natexlab{}.
\newblock \showarticletitle{Response time and display rate in human performance
  with computers}.
\newblock \bibinfo{journal}{\emph{ACM Computing Surveys (CSUR)}}
  \bibinfo{volume}{16}, \bibinfo{number}{3} (\bibinfo{year}{1984}),
  \bibinfo{pages}{265--285}.
\newblock


\bibitem[\protect\citeauthoryear{Staffa and Rossi}{Staffa and Rossi}{2016}]%
        {staffa2016recommender}
\bibfield{author}{\bibinfo{person}{Mariacarla Staffa} {and}
  \bibinfo{person}{Silvia Rossi}.} \bibinfo{year}{2016}\natexlab{}.
\newblock \showarticletitle{Recommender interfaces: the more human-like, the
  more humans like}. In \bibinfo{booktitle}{\emph{Social Robotics: 8th
  International Conference, ICSR 2016, Kansas City, MO, USA, November 1-3, 2016
  Proceedings 8}}. Springer, \bibinfo{pages}{200--210}.
\newblock


\bibitem[\protect\citeauthoryear{Stupak}{Stupak}{2009}]%
        {stupak2009time}
\bibfield{author}{\bibinfo{person}{Noah Stupak}.}
  \bibinfo{year}{2009}\natexlab{}.
\newblock \bibinfo{booktitle}{\emph{Time delays and system response times in
  human-computer interaction}}.
\newblock \bibinfo{publisher}{Rochester Institute of Technology}.
\newblock


\bibitem[\protect\citeauthoryear{Tanaka}{Tanaka}{2023}]%
        {tanaka20234}
\bibfield{author}{\bibinfo{person}{Takahiro Tanaka}.}
  \bibinfo{year}{2023}\natexlab{}.
\newblock \showarticletitle{4 Robotic Human--Machine Interface Towards Driving
  Behavior Improvement for Elderly Drivers}.
\newblock \bibinfo{journal}{\emph{Towards Human-Vehicle Harmonization}}
  \bibinfo{volume}{3} (\bibinfo{year}{2023}), \bibinfo{pages}{47}.
\newblock


\bibitem[\protect\citeauthoryear{Tanaka, Fujikake, Yoshihara, Karatas,
  Shimazaki, Kanamori, and Aoki}{Tanaka et~al\mbox{.}}{2020}]%
        {tanaka2020analysis}
\bibfield{author}{\bibinfo{person}{Takahiro Tanaka}, \bibinfo{person}{Kazuhiro
  Fujikake}, \bibinfo{person}{Yuki Yoshihara}, \bibinfo{person}{Nihan Karatas},
  \bibinfo{person}{Kan Shimazaki}, \bibinfo{person}{Hitoshi Kanamori}, {and}
  \bibinfo{person}{Hirofumi Aoki}.} \bibinfo{year}{2020}\natexlab{}.
\newblock \showarticletitle{Analysis of Distraction and Driving Behavior
  Improvement Using a Driving Support Agent for Elderly and Non-Elderly Drivers
  on Public Roads}. In \bibinfo{booktitle}{\emph{2020 IEEE Intelligent Vehicles
  Symposium (IV)}}. IEEE, \bibinfo{pages}{1029--1034}.
\newblock


\bibitem[\protect\citeauthoryear{Tashev, Seltzer, Ju, Wang, and Acero}{Tashev
  et~al\mbox{.}}{2009}]%
        {tashev2009commute}
\bibfield{author}{\bibinfo{person}{Ivan Tashev}, \bibinfo{person}{Mike
  Seltzer}, \bibinfo{person}{Yun-Cheng Ju}, \bibinfo{person}{Ye-Yi Wang}, {and}
  \bibinfo{person}{Alex Acero}.} \bibinfo{year}{2009}\natexlab{}.
\newblock \showarticletitle{Commute UX: Voice enabled in-car infotainment
  system}.
\newblock  (\bibinfo{year}{2009}).
\newblock


\bibitem[\protect\citeauthoryear{Thellman, Hagman, Jonsson, Nilsson,
  Samuelsson, Simonsson, Sk{\"o}nvall, Westin, and Silvervarg}{Thellman
  et~al\mbox{.}}{2018}]%
        {thellman2018he}
\bibfield{author}{\bibinfo{person}{Sam Thellman}, \bibinfo{person}{William
  Hagman}, \bibinfo{person}{Emma Jonsson}, \bibinfo{person}{Lisa Nilsson},
  \bibinfo{person}{Emma Samuelsson}, \bibinfo{person}{Charlie Simonsson},
  \bibinfo{person}{Julia Sk{\"o}nvall}, \bibinfo{person}{Anna Westin}, {and}
  \bibinfo{person}{Annika Silvervarg}.} \bibinfo{year}{2018}\natexlab{}.
\newblock \showarticletitle{He is not more persuasive than her: No gender
  biases toward robots giving speeches}. In
  \bibinfo{booktitle}{\emph{Proceedings of the 18th International Conference on
  Intelligent Virtual Agents}}. \bibinfo{pages}{327--328}.
\newblock


\bibitem[\protect\citeauthoryear{Van~Compernolle, Ma, Xie, and
  Van~Diest}{Van~Compernolle et~al\mbox{.}}{1990}]%
        {van1990speech}
\bibfield{author}{\bibinfo{person}{Dirk Van~Compernolle},
  \bibinfo{person}{Weiye Ma}, \bibinfo{person}{Fei Xie}, {and}
  \bibinfo{person}{Marc Van~Diest}.} \bibinfo{year}{1990}\natexlab{}.
\newblock \showarticletitle{Speech recognition in noisy environments with the
  aid of microphone arrays}.
\newblock \bibinfo{journal}{\emph{Speech Communication}} \bibinfo{volume}{9},
  \bibinfo{number}{5-6} (\bibinfo{year}{1990}), \bibinfo{pages}{433--442}.
\newblock


\bibitem[\protect\citeauthoryear{Van Der~Laan, Heino, and De~Waard}{Van
  Der~Laan et~al\mbox{.}}{1997}]%
        {van1997simple}
\bibfield{author}{\bibinfo{person}{Jinke~D Van Der~Laan},
  \bibinfo{person}{Adriaan Heino}, {and} \bibinfo{person}{Dick De~Waard}.}
  \bibinfo{year}{1997}\natexlab{}.
\newblock \showarticletitle{A simple procedure for the assessment of acceptance
  of advanced transport telematics}.
\newblock \bibinfo{journal}{\emph{Transportation Research Part C: Emerging
  Technologies}} \bibinfo{volume}{5}, \bibinfo{number}{1}
  (\bibinfo{year}{1997}), \bibinfo{pages}{1--10}.
\newblock


\bibitem[\protect\citeauthoryear{Wang, Lee, Kamalesh~Sanghavi, Eskew, Zhou, and
  Jeon}{Wang et~al\mbox{.}}{2021}]%
        {wang2021vehicle}
\bibfield{author}{\bibinfo{person}{Manhua Wang}, \bibinfo{person}{Seul~Chan
  Lee}, \bibinfo{person}{Harsh Kamalesh~Sanghavi}, \bibinfo{person}{Megan
  Eskew}, \bibinfo{person}{Bo Zhou}, {and} \bibinfo{person}{Myounghoon Jeon}.}
  \bibinfo{year}{2021}\natexlab{}.
\newblock \showarticletitle{In-vehicle intelligent agents in fully autonomous
  driving: The effects of speech style and embodiment together and separately}.
  In \bibinfo{booktitle}{\emph{13th International Conference on Automotive User
  Interfaces and Interactive Vehicular Applications}}.
  \bibinfo{pages}{247--254}.
\newblock


\bibitem[\protect\citeauthoryear{Wang, Li, Bell, and Lai}{Wang
  et~al\mbox{.}}{2023}]%
        {wang2023cross}
\bibfield{author}{\bibinfo{person}{Yaoting Wang}, \bibinfo{person}{Yuanchao
  Li}, \bibinfo{person}{Peter Bell}, {and} \bibinfo{person}{Catherine Lai}.}
  \bibinfo{year}{2023}\natexlab{}.
\newblock \showarticletitle{Cross-Attention is Not Enough: Incongruity-Aware
  Multimodal Sentiment Analysis and Emotion Recognition}.
\newblock \bibinfo{journal}{\emph{arXiv preprint arXiv:2305.13583}}
  (\bibinfo{year}{2023}).
\newblock


\bibitem[\protect\citeauthoryear{Wang, Acero, and Chelba}{Wang
  et~al\mbox{.}}{2003}]%
        {wang2003word}
\bibfield{author}{\bibinfo{person}{Ye-Yi Wang}, \bibinfo{person}{Alex Acero},
  {and} \bibinfo{person}{Ciprian Chelba}.} \bibinfo{year}{2003}\natexlab{}.
\newblock \showarticletitle{Is word error rate a good indicator for spoken
  language understanding accuracy}. In \bibinfo{booktitle}{\emph{2003 IEEE
  workshop on automatic speech recognition and understanding (IEEE Cat. No.
  03EX721)}}. IEEE, \bibinfo{pages}{577--582}.
\newblock


\bibitem[\protect\citeauthoryear{Ward, Li, Zhao, and Kawahara}{Ward
  et~al\mbox{.}}{2016}]%
        {ward2016interactional}
\bibfield{author}{\bibinfo{person}{Nigel~G Ward}, \bibinfo{person}{Yuanchao
  Li}, \bibinfo{person}{Tianyu Zhao}, {and} \bibinfo{person}{Tatsuya
  Kawahara}.} \bibinfo{year}{2016}\natexlab{}.
\newblock \showarticletitle{Interactional and pragmatics-related prosodic
  patterns in Mandarin dialog}. In \bibinfo{booktitle}{\emph{Speech prosody}}.
\newblock


\bibitem[\protect\citeauthoryear{Williams, Peters, and Breazeal}{Williams
  et~al\mbox{.}}{2013}]%
        {williams2013towards}
\bibfield{author}{\bibinfo{person}{Kenton~J Williams},
  \bibinfo{person}{Joshua~C Peters}, {and} \bibinfo{person}{Cynthia~L
  Breazeal}.} \bibinfo{year}{2013}\natexlab{}.
\newblock \showarticletitle{Towards leveraging the driver's mobile device for
  an intelligent, sociable in-car robotic assistant}. In
  \bibinfo{booktitle}{\emph{2013 IEEE intelligent vehicles symposium (IV)}}.
  IEEE, \bibinfo{pages}{369--376}.
\newblock


\bibitem[\protect\citeauthoryear{Wu, Gorospe, Misu, Huynh, and Guerrero}{Wu
  et~al\mbox{.}}{2020}]%
        {wu2020driving}
\bibfield{author}{\bibinfo{person}{Xingwei Wu}, \bibinfo{person}{Yuki Gorospe},
  \bibinfo{person}{Teruhisa Misu}, \bibinfo{person}{Y Huynh}, {and}
  \bibinfo{person}{Nimsi Guerrero}.} \bibinfo{year}{2020}\natexlab{}.
\newblock \showarticletitle{What driving says about you: A small-sample
  exploratory study between personality and self-reported driving style among
  young male drivers}. In \bibinfo{booktitle}{\emph{12th International
  Conference on Automotive User Interfaces and Interactive Vehicular
  Applications}}. \bibinfo{pages}{104--110}.
\newblock


\bibitem[\protect\citeauthoryear{Xia and Hafner}{Xia and Hafner}{2021}]%
        {xia2021engaging}
\bibfield{author}{\bibinfo{person}{Sichen~Ada Xia} {and}
  \bibinfo{person}{Christoph~A Hafner}.} \bibinfo{year}{2021}\natexlab{}.
\newblock \showarticletitle{Engaging the online audience in the digital era: A
  multimodal analysis of engagement strategies in TED talk videos}.
\newblock \bibinfo{journal}{\emph{Ib{\'e}rica}} \bibinfo{number}{42}
  (\bibinfo{year}{2021}), \bibinfo{pages}{33--58}.
\newblock


\bibitem[\protect\citeauthoryear{Yamada, Shinkawa, Kobayashi, Takagi, Nemoto,
  Nemoto, and Arai}{Yamada et~al\mbox{.}}{2021}]%
        {yamada2021using}
\bibfield{author}{\bibinfo{person}{Yasunori Yamada}, \bibinfo{person}{Kaoru
  Shinkawa}, \bibinfo{person}{Masatomo Kobayashi}, \bibinfo{person}{Hironobu
  Takagi}, \bibinfo{person}{Miyuki Nemoto}, \bibinfo{person}{Kiyotaka Nemoto},
  {and} \bibinfo{person}{Tetsuaki Arai}.} \bibinfo{year}{2021}\natexlab{}.
\newblock \showarticletitle{Using speech data from interactions with a voice
  assistant to predict the risk of future accidents for older drivers:
  prospective cohort study}.
\newblock \bibinfo{journal}{\emph{Journal of medical internet research}}
  \bibinfo{volume}{23}, \bibinfo{number}{4} (\bibinfo{year}{2021}),
  \bibinfo{pages}{e27667}.
\newblock


\bibitem[\protect\citeauthoryear{Yamaguchi, Inoue, Yoshino, Takanashi, Ward,
  and Kawahara}{Yamaguchi et~al\mbox{.}}{2016}]%
        {yamaguchi2016analysis}
\bibfield{author}{\bibinfo{person}{Takashi Yamaguchi}, \bibinfo{person}{Koji
  Inoue}, \bibinfo{person}{Koichiro Yoshino}, \bibinfo{person}{Katsuya
  Takanashi}, \bibinfo{person}{Nigel~G Ward}, {and} \bibinfo{person}{Tatsuya
  Kawahara}.} \bibinfo{year}{2016}\natexlab{}.
\newblock \showarticletitle{Analysis and prediction of morphological patterns
  of backchannels for attentive listening agents}. In
  \bibinfo{booktitle}{\emph{Proc. 7th International Workshop on Spoken Dialogue
  Systems}}. \bibinfo{pages}{1--12}.
\newblock


\bibitem[\protect\citeauthoryear{Yuan, Thompson, Watson, Chase, Senthilkumar,
  Brush, and Yarosh}{Yuan et~al\mbox{.}}{2019}]%
        {yuan2019speech}
\bibfield{author}{\bibinfo{person}{Ye Yuan}, \bibinfo{person}{Stryker
  Thompson}, \bibinfo{person}{Kathleen Watson}, \bibinfo{person}{Alice Chase},
  \bibinfo{person}{Ashwin Senthilkumar}, \bibinfo{person}{AJ~Bernheim Brush},
  {and} \bibinfo{person}{Svetlana Yarosh}.} \bibinfo{year}{2019}\natexlab{}.
\newblock \showarticletitle{Speech interface reformulations and voice assistant
  personification preferences of children and parents}.
\newblock \bibinfo{journal}{\emph{International Journal of Child-Computer
  Interaction}}  \bibinfo{volume}{21} (\bibinfo{year}{2019}),
  \bibinfo{pages}{77--88}.
\newblock


\bibitem[\protect\citeauthoryear{Zargham, Alexandrovsky, Erich, Wenig, and
  Malaka}{Zargham et~al\mbox{.}}{2022}]%
        {zargham2022want}
\bibfield{author}{\bibinfo{person}{Nima Zargham}, \bibinfo{person}{Dmitry
  Alexandrovsky}, \bibinfo{person}{Jan Erich}, \bibinfo{person}{Nina Wenig},
  {and} \bibinfo{person}{Rainer Malaka}.} \bibinfo{year}{2022}\natexlab{}.
\newblock \showarticletitle{“I Want It That Way”: Exploring Users’
  Customization and Personalization Preferences for Home Assistants}. In
  \bibinfo{booktitle}{\emph{CHI Conference on Human Factors in Computing
  Systems Extended Abstracts}}. \bibinfo{pages}{1--8}.
\newblock


\bibitem[\protect\citeauthoryear{Zepf, Dittrich, Hernandez, and Schmitt}{Zepf
  et~al\mbox{.}}{2019}]%
        {zepf2019towards}
\bibfield{author}{\bibinfo{person}{Sebastian Zepf}, \bibinfo{person}{Monique
  Dittrich}, \bibinfo{person}{Javier Hernandez}, {and}
  \bibinfo{person}{Alexander Schmitt}.} \bibinfo{year}{2019}\natexlab{}.
\newblock \showarticletitle{Towards empathetic car interfaces: Emotional
  triggers while driving}. In \bibinfo{booktitle}{\emph{Extended Abstracts of
  the 2019 CHI Conference on Human Factors in Computing Systems}}.
  \bibinfo{pages}{1--6}.
\newblock


\bibitem[\protect\citeauthoryear{Zihsler, Hock, Walch, Dzuba, Schwager, Szauer,
  and Rukzio}{Zihsler et~al\mbox{.}}{2016}]%
        {zihsler2016carvatar}
\bibfield{author}{\bibinfo{person}{Jens Zihsler}, \bibinfo{person}{Philipp
  Hock}, \bibinfo{person}{Marcel Walch}, \bibinfo{person}{Kirill Dzuba},
  \bibinfo{person}{Denis Schwager}, \bibinfo{person}{Patrick Szauer}, {and}
  \bibinfo{person}{Enrico Rukzio}.} \bibinfo{year}{2016}\natexlab{}.
\newblock \showarticletitle{Carvatar: increasing trust in highly-automated
  driving through social cues}. In \bibinfo{booktitle}{\emph{Adjunct
  proceedings of the 8th international conference on automotive user interfaces
  and interactive vehicular applications}}. \bibinfo{pages}{9--14}.
\newblock


\end{thebibliography}

\end{document}